\documentclass[12pt, hyper]{article}
\pdfoutput=1
\usepackage{amsmath,amsopn,amssymb,amsfonts}
\usepackage{mathtools}
\usepackage{amsthm}
\usepackage{hyperref}
\usepackage{lscape}
\usepackage[pdftex]{graphicx}
\usepackage{epsfig}
\usepackage{mathrsfs}
\DeclareMathAlphabet{\mathpzc}{OT1}{pzc}{m}{it}
\usepackage{verbatim}
\usepackage{multirow}
\usepackage{amsmath,amsfonts,amsthm,bm}
\usepackage{pgfplots}
\usepackage{pgf}
\usepackage{tikz}
\usetikzlibrary{arrows,automata}
\usepackage[latin1]{inputenc}
\usepackage{verbatim}
\usetikzlibrary{arrows}
\usepackage{enumitem}
\usepackage{arydshln}
\usepackage{cite}
\usepackage[absolute,overlay]{textpos}
\usepackage[margin=1cm]{caption}
\usepackage[font={small}]{caption}

\def\CD{{\cal D}}

\def\CM{{\cal M}}
\def\CN{{\cal N}}

\def\e{\epsilon}

\def\l{\lambda}

\def\s{\sigma}
\def\t{\tau}

\def\D{\Delta}

\usepackage{cleveref}

\usetikzlibrary{arrows,positioning}
\tikzset{
    >=stealth',
    punkt/.style={
           rectangle,
           rounded corners,
           draw=black, very thick,
           text width=6.5em,
           minimum height=2em,
           text centered},
    pil/.style={
           ->,
           thick,
           shorten <=2pt,
           shorten >=2pt,}
}
\numberwithin{equation}{section}
\begin{document}

\begin{titlepage}
\vfill
\begin{flushright}
{\tt\normalsize KIAS-P18107}\\
\end{flushright}
\vfill
\begin{center}
{\LARGE\bf Monster Anatomy}

\vfill

Jin-Beom Bae, Kimyeong Lee and Sungjay Lee

\vskip 5mm
{\it Korea Institute for Advanced Study \\
85 Hoegiro, Dongdaemun-Gu, Seoul 02455, Korea}

\end{center}
\vfill

\begin{abstract}
\noindent We investigate the two-dimensional conformal field theories (CFTs) of $c=\frac{47}{2}$, $c=\frac{116}{5}$ and $c=23$ 
`dual' to the critical Ising model, the three state Potts model and the tensor product of two Ising models, respectively. We argue that these CFTs exhibit moonshines for the double covering of the baby Monster group, $2\cdot \mathbb{B}$, the triple covering of the largest Fischer group,
$3\cdot \text{Fi}_{24}'$ and multiple-covering of the second largest Conway group, 
$2\cdot 2^{1+22} \cdot \text{Co}_2$. 
Various twined characters are shown to satisfy generalized bilinear relations involving 
Mckay-Thompson series.
We also rediscover that the `self-dual' two-dimensional bosonic conformal field theory of $c=12$ has the Conway group $\text{Co}_{0}\simeq2\cdot\text{Co}_1$ as an automorphism group.
\end{abstract}

\vfill
\end{titlepage}

\parskip 0.1 cm
\renewcommand{\thefootnote}{\#\arabic{footnote}}
\setcounter{footnote}{0}

\parskip 0.2 cm

\section{Introduction}
Mckay and Thompson's remarkable observation between the monster group $\mathbb M$ and the  modular objects, especially, the $ j$-invariant, motivated the study of the so-called 'Monstrous Moonshine' in \cite{Conway1979}. Each Fourier coefficient of the modular invariant $j(\t)-744$ ($q=e^{2\pi i\t}$), 
which can describe the partition function of $c=24$ chiral CFT,
\begin{align}
\label{j-func}
  j(\t) -744= \frac{1}{q}  + 196884 q + 21493760 q^2 + \cdots,
\end{align}
can be decomposed into the dimension of the irreducible representation of the monster group $\mathbb{M}$.
Frenkel, Lepowsky and Meurman\cite{Frenkel1988} provided a (heuristic) derviation
of the Monstrous moonshine from an explicit construction
of the chiral CFT based on the Leech lattice followed by a
$\mathbb{Z}_2$ orbifold. Many examples of the generalizations of moonshine phenomena with 
different sporadic groups have been uncovered in the last 
decades\cite{Duncan2005,Eguchi2011a,Cheng2013,Cheng2014,Cheng2014a,Duncan2014a,Cheng2015,Harvey2016,Duncan2017}.

In this article, we utilize a holomorphic bilinear relation\cite{Hampapura2016} to further
explore a new class of moonshine phenomena. It has been observed recently
that characters $f_i(\t)$ of a certain rational CFT with central charge $c$
obey an intriguing bilinear relation giving a modular invariant $j(\t)$,
\begin{align}
  \label{bilienar relation}
  \sum_{i=0}^{n-1} f_i(\t) \tilde f_i(\t) = j(\t) - 744,
\end{align}
where $\tilde f_i(\t)$ can be interpreted as characters of a `dual' rational
CFT with central charge $(24-c)$. For instance, the critical
Ising model with $c=\frac{1}{2}$ and a rational CFT with $c=\frac{47}{2}$ satisfies
the bilinear relation. Another example is a pair of
rational CFTs of $c=8$ and $c=16$ having no Kac-Moody symmetry but finite
group symmetry\cite{Bae2017}. Further examples can be found in \cite{Chandra2018}.

The rational CFT with $c=\frac{47}{2}$ dual to the critical Ising model
exhibits Moonshine for the baby Monster group, second largest sporadic
group. It is challenging to search for a dual rational CFT
showing Moonshine for the sporadic groups other than
the baby Monster group. The search first requires
an explicit $q$-expansion of each character in two
rational CFTs of dual pair. To do so, we make use of
a modular-invariant differential equation (MDE) of the form below\cite{Mathur1988}
\begin{align}
  \left[ \CD_\t^n + \sum_{k=0}^{n-1} \phi_{2(n-k)}(\t) \CD_\t^k \right] f(\t) = 0,
  \label{MDE01}
\end{align}
where $\CD_\t$ denotes the Serre derivative acting on
a modular form of weight $r$,
\begin{align}
   \CD_\tau\equiv \partial_\t - \frac16 i\pi r E_2(\t),
\end{align}
and $\phi_k(\t)$ are modular forms of weight $k$.
The MDE can be used to explore the space of rational
CFTs. This is because solutions to an MDE,
which furnish a finite-dimensional representation of $SL(2,\mathbb{Z})$,
can play a role as candidate characters $f_i(\t)$ ($i=0,1,2,..,n-1$)
in a rational CFT. One can show from (\ref{MDE01}) that the
conformal weights $h_i$ of primaries and the central charge $c$ of
a candidate RCFT have to satisfy the relation below
\begin{align}
\label{condition01}
  \sum_{i=0}^{n-1} \Big[ h_i - \frac{c}{24} \Big] =
  \frac{n(n-1)}{12} - \frac{l}{6},
\end{align}
where $l$ is a non-negative integer other than $1$.
(\ref{condition01}) implies that a rational CFT can be
characterized by conformal weights $h_i$ and a number $n$ of primaries, the central
charge $c$ and an integer $l$.
When $(\{h_i\},n,c,l)$  of a rational CFT
are related to  $(\{\tilde h_i\},n,\tilde c,\tilde l)$ of another rational CFT as follows
\begin{align}
  h_0=\tilde h_0 = 0, \qquad h_i + \tilde h_i = 2 \ \text{ for } \ i\neq 0,
\end{align}
and
\begin{align}
  c + \tilde c =24, \qquad l+\tilde l = (n-3)(n-4),
\end{align}
the characters of two rational CFTs can obey the bilinear
relation (\ref{bilienar relation}). Namely, one rational CFT is dual to the other.

\begin{figure}[t!]
\begin{center}
\begin{tikzpicture}[node distance=1cm]
 \node[punkt] (Monster) {$\mathbb{M}$};
 \node[punkt, below=1cm of Monster] (BabyMonster) {2$ \cdot \mathbb{B}$};
 \node[punkt, below right=1.414cm of Monster] (Fi24) {3 $ \cdot$ Fi$_{24}'$};
 \node[punkt, below left=1.414cm of Monster] (Co1) {$2^{1+24}$ $ \cdot$ Co$_1$};
 \node[punkt, below =1cm of Fi24] (Fi23) {Fi$_{23}$};
 \node[punkt, below =1cm of Co1] (Co2) {$2 \cdot 2^{1+22} \cdot$ Co$_2$};
 \node[punkt, below =2cm of BabyMonster] (M24) {$2^{12} \cdot$ M$_{24}$};
 \draw [draw=red,->] (Monster) -- (BabyMonster) ;
 \draw [draw=blue,->] (Monster) -- (Fi24) ;
 \draw [dashed, ->] (Monster) -- (Co1) ;
 \draw [->] (Fi24) -- (Fi23) ;
 \draw [->] (Co1) -- (Co2) ;
 \draw [->] (BabyMonster) -- (Fi23) ;
 \draw [draw=red,->] (BabyMonster) -- (Co2) ;
 \draw [->] (Co1) -- (M24) ;
 \draw [draw=green,->] (Fi24) -- (M24) ;
\end{tikzpicture}
\end{center}
\caption{Partial flows of maximal subgroups from the monster group 
$\mathbb{M}$\cite{Wilson2017}. Each arrow
from A to B implies that B is a maximal subgroup of A.}
\label{Monster Decomposition}
\end{figure}
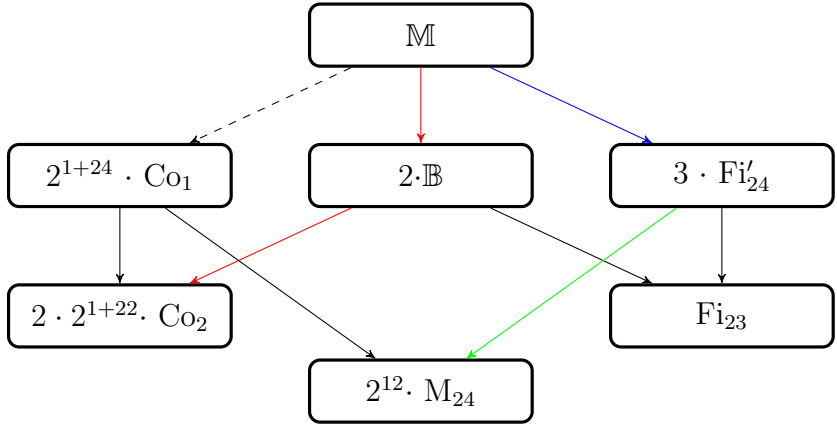
We analyze rational CFTs dual to the three-state Potts model and to
the product of two Ising models. Interestingly enough, our results can propose that
the dual rational CFTs show novel connections to the triple covering 
of the largest Fischer group, $3\cdot \text{Fi}_{24}'$ and the 
multiple-covering of the second largest Conway group, $2\cdot 2^{1+22}\cdot \text{Co}_2$, respectively.
Note that $3 \cdot \text{Fi}_{24}'$ is a maximal subgroup of $\mathbb{M}$ described in 
Figure \ref{Monster Decomposition}.
We also observe that the self-dual theory of $c=12$ has the largest
Conway group Co$_0$ as an automorphism group. In fact, we can show
that the self-dual theory can be identified as the GSO projection of
the well-known $\CN=1$ superconformal extremal CFT of $c=12$\cite{Frenkel1988,Witten2007}.
Our observations suggest that the subgroup decomposition along the red and the blue arrows 
in Figure \ref{Monster Decomposition} can be realized by the holomorphic bilinear relation.
The moonshines for $2\cdot \mathbb{B}$ in baby Monster CFT and $3\cdot \text{Fi}_{24}'$ 
in $c=\frac{116}{5}$ CFT are further supported by generalized bilinear relations involving Mckay-Thompson series. 
The details will be shown in section \ref{sec:2} and section \ref{sec:3}.

\section{Dual of the Ising Model and the Baby Monster}
\label{sec:2}

The simplest unitary minimal model $\CM(4,3)$
describes the critical point of the second-order
phase transition of the Ising model.
The critical Ising model has the identity ${\bf 1}$,
the energy density $\e$ and  the spin $\s$ whose
scaling dimensions are $\D=0,1,\frac{1}{8}$ respectively.
Assuming three operators have no spin, one can
relate them with allowed primary
fields $\phi_{1,1}, \phi_{2,1}, \phi_{2,2}$ of weight $h=0,\frac{1}{2},\frac{1}{16}$ in $\CM(4,3)$ as follows
\begin{align}
\begin{split}
  {\bf 1} \leftrightarrow \phi_{1,1} \\
  \e \leftrightarrow \phi_{2,1} \\
  \s \leftrightarrow \phi_{2,2}
\end{split}
\end{align}
The $c=\frac{1}{2}$ Virasoro characters for three primary fields
are
\begin{align}
\label{Ising Ch}
\begin{split}
  f_0(\t) & = \frac{1}{2} \Big( \sqrt{\frac{\vartheta_3(\tau)}{\eta(\tau)}}
  + \sqrt{\frac{\vartheta_4(\tau)}{\eta(\tau)}} \Big),
  \\
  f_\e(\t) & = \frac{1}{2} \Big( \sqrt{\frac{\vartheta_3(\tau)}{\eta(\tau)}} - \sqrt{\frac{\vartheta_4(\tau)}{\eta(\tau)}} \Big) ,
  \\
  f_\s(\t) & =  \frac{1}{\sqrt{2}} \sqrt{\frac{\vartheta_2(\tau)}{\eta(\tau)}}.
\end{split}
\end{align}
These characters transform into one another under $SL(2,\mathbb{Z})$.
In particular, the modular S-matrix is
\begin{align}
\label{Ising S-law}
\begin{pmatrix}
f_0(-1/{\tau}) \\ f_\e(-1/{\tau}) \\ f_\s(-1/{\tau})
\end{pmatrix} = \frac{1}{2}
\begin{pmatrix}
1 & 1 & \sqrt{2} \\
1 & 1 & -\sqrt{2}  \\
\sqrt{2} & -\sqrt{2} & 0
\end{pmatrix}
\begin{pmatrix} f_0(\t) \\ f_\e(\t) \\ f_\s(\t) \end{pmatrix} .
\end{align}

It has been shown recently in \cite{Hampapura2016} that the characters of
the critical Ising model obey an intriguing bilinear relation,
\begin{align}
\label{Ising BM}
  j(\tau)-744 = f_0(\tau)  \cdot \tilde f_0(\t) + f_\e(\t) \cdot \tilde f_\e(\tau)
  + f_\s(\t) \cdot \tilde f_\s(\tau),
\end{align}
where $j(\t)-744$ is the Monster module,
\begin{align}
  j(\t) = \frac{12^3 E_4^3(\t)}{E_4^3(\t) - E_6^2(\t)} = \frac{1}{q} + 744 + 196884 q + 21493760 q^2 + \cdots.
\end{align}
Here $\tilde f_0$, $\tilde f_\s$ and $\tilde f_\e$ are three independent solutions
to a third order modular differential equation
\begin{align}
\begin{split}
0 = \Big[ \CD_\tau^3 + \frac{2315 \pi^2}{576} E_4(\tau) \CD_\tau - i  \frac{27025 \pi^3}{6912} E_6(\tau) \Big] \tilde{f},
\end{split}
\end{align}
and can be expanded in powers of $q$ as follow.
\begin{align}
\label{BM characters}
\begin{split}
  \tilde f_{0}(\tau) &=
  q^{-\frac{47}{48}}\left(1 +96256' q^2 +9646891 q^3 + 366845011 q^4 + \cdots\right),
  \\
  \tilde f_\e(\tau) &=
  q^{\frac{25}{48}}\left(4371 + 1143745 q + 64680601 q^2 + 1829005611 q^3 + \cdots\right),
  \\
  \tilde f_\s(\tau) &=
  q^{\frac{23}{24}}\left(96256 + 10602496 q + 420831232 q^2 + 9685952512 q^3+ \cdots\right).
\end{split}
\end{align}
In fact, (\ref{BM characters}) can be identified as
three characters of conformal weights $h=0,\frac{31}{16},\frac{3}{2}$ in the dual RCFT with $c=\frac{47}{2}$.


Since the $j(\t)$ is invariant under $SL(2,\mathbb{Z})$,
the modular S-matrix of the dual RCFT with $c=\frac{47}{2}$,
often referred to as the baby Monster CFT, obtained from that of 
the critical Ising model, namely
\begin{align}
\label{BM S-law}
\begin{pmatrix}
\tilde f_0(-1/{\tau}) \\ \tilde f_\e(-1/{\tau}) \\ \tilde f_\s(-1/{\tau})
\end{pmatrix} = \frac{1}{2}
\begin{pmatrix}
1 & 1 & \sqrt{2} \\
1 & 1 & -\sqrt{2}  \\
\sqrt{2} & -\sqrt{2} & 0
\end{pmatrix}
\begin{pmatrix} \tilde f_0(\t) \\ \tilde f_\e(\t) \\ \tilde f_\s(\t) \end{pmatrix} .
\end{align}
In addition, the Verlinde formulae implies that these two RCFTs also share the same
fusion rule algebra. From \eqref{BM S-law}, the modular invariant torus partition function of
the baby Monster CFT has to be
\begin{align}
\label{BM CFT partition function}
  Z_{c=\frac{47}{2}} (\tau, \bar{\tau}) =  \tilde f_{0}(\tau) \bar{\tilde f}_{0}(\bar{\tau})
  +  \tilde f_\e(\tau) \bar{\tilde f}_{\e}(\bar{\tau})
  +  \tilde f_{\s}(\tau)  \bar{\tilde f}_{\s}(\bar{\tau}),
\end{align}
which agrees with numerical results in \cite{Bae2017}.

It is known that dual RCFT with $c=\frac{47}{2}$ has baby Monster group $\mathbb{B}$ 
as a finite group symmetry\cite{Hampapura2016}, because \eqref{BM characters} are identical to the characters 
of modules $V\mathbb{B}^{\natural}_{(0)}, V\mathbb{B}^{\natural}_{(1)}$ and $V\mathbb{B}^{\natural}_{(2)}$ in \cite{Hoehn2007}, respectively. In this paper, we conjecture that the  
finite group symmetry of $c=\frac{47}{2}$ RCFT can be promoted to the double covering of baby Monster 
group $2\cdot \mathbb{B}$.
As a demonstration, one can show that each coefficient in (\ref{BM characters}) can be expressed as
a sum of dimensions of irreducible representations of $2\cdot\mathbb{B}$, e.g.
\begin{align}
\label{Decomposition01}
\begin{split}
  &4371  = {\bf 4371}, \quad  96256'  = {\bf 1} \oplus {\bf 96255}, \quad  96256 = {\bf 96256}, \\
  &9646891 =   {\bf{1}} \oplus {\bf{96255}} \oplus {\bf{9550635}}, \quad 1143745 = {\bf{4371}} \oplus {\bf{1139374}}, \\
& 10602496 = {\bf{96256}}  \oplus {\bf{10506240}}, \\ 
  &366845011  = 2 \cdot {\bf 1} \oplus  2 \cdot  {\bf{96255}} \oplus {\bf{9458750}} \oplus {\bf{9550635}} \oplus {\bf{347643114}}, \\
  &64680601  =  2 \cdot {\bf 4371} \oplus {\bf{1139374}} \oplus {\bf{63532485}}, \\
  &420831232  =  2 \cdot {\bf 96256}  \oplus {\bf{10506240}} \oplus {\bf{410132480}}.
\end{split}
\end{align}
Note that ${\bf 1}$ in the second line of (\ref{Decomposition01})
can be identified as the Virasoro descendent of the vacuum.

We apply a refined test proposed in \cite{Gaberdiel2010,Eguchi2010} to provide further supporting evidence that $c=\frac{47}{2}$ CFT 
exhibit moonshine for $2\cdot \mathbb{B}$. To this end, we introduce the twined character,
\begin{align}
\begin{split}
\label{Twisted PF}
\tilde{f}^{g}_{i}(\tau)=\text{Tr}_{\mathcal{H}_i} \left[g\cdot q^{h-\frac{c}{24}}\right].
\end{split}
\end{align}
where  $g$ is a group element of $ 2 \cdot \mathbb{B}$ and the trace is taken over all states in the Hilbert space $\mathcal{H}_i$,
that consist of primary state of weight $h_i$ and its descendants.
As illustrated in \cite{Gaberdiel2010}, it is straightforward to obtain the twined characters using the 
character table of $2\cdot\mathbb{B}$ in appendix \ref{App:B}. 
For instance, let us consider the twined characters for $g = 2C$.
Explicitly, the first term in $\tilde f_\s(\tau)$, {\bf{96256}}, 
is replaced by {\bf{2048}}. In this way, one can check that the twined characters are given by
\begin{align}
\begin{split}
\label{2C twisted}
\tilde{f}_0^{2C}(q) &= q^{-\frac{47}{48}} \left( 1 + 2048 q^2 + 37675 q^3 + 470099 q^4 + \cdots \right), \\
\tilde{f}_{\epsilon}^{2C}(q) &= q^{\frac{25}{48}} \left( 275 + 9153 q + 144025 q^2 + \cdots \right),  \\
\tilde{f}_{\sigma}^{2C}(q) &= q^{\frac{23}{24}} \left( 2048 + 47104 q + 565248 q^2 + \cdots \right). \\
\end{split}
\end{align}
Intriguingly, we find that \eqref{2C twisted} satisfy a bilinear relation of the form
\begin{align}
\begin{split}
\label{extended bilinear}
  j^{2A}(\tau) &= f_0(\tau) \cdot \tilde f^{2C}_0(\t) + f_\e(\t) \cdot \tilde f^{2C}_\e(\tau) + f_\s(\t) \cdot \tilde f^{2C}_\s(\tau),
\end{split}
\end{align}
where $j^{2A}(\tau)$ is Mckay-Thompson series of class 2A,
\begin{align}
\begin{split}
j^{2A}(\tau)  &= \frac{\eta(q)^{24}}{\eta(q^{2})^{24}} +  \frac{2^{12} \eta(q^{2})^{24}}{\eta(q)^{24}} + 24 \\
              &= \frac{1}{q} + 4372 q + 96256 q^2 + 1240002 q^3 +10698752 q^4+ \cdots. \\
\end{split}
\end{align}
In appendix \ref{App:C}, we present list of the generalized bilinear relations for various $g \in 2 \cdot \mathbb{B}$.
Combined with the characters of the Ising model, all the twined characters we investigated constitutes Mckay-Thompson series of certain class.  
Sometimes, combination of the twined characters for different group elements and characters of the critical Ising model 
yields Mckay-Thompson series of identical class. More precisely, even if we replace 2C characters to 2A(or 2B) characters in 
the right-hand side of \eqref{extended bilinear}, the left-hand side of \eqref{extended bilinear} still remained as $j^{2A}(\tau)$. Note that the decompositions in \eqref{Decomposition01} are consistent to bilinear relations.

\section{Dual of the Three-State Potts Model and $3\cdot\text{Fi}_{24}'$}
\label{sec:3}

We now make use of the bilinear relations but with different pairs of
characters to look for new RCFTs related to the other sporadic groups.
In particular, we propose in this section that a dual CFT of the three-state Potts model
has the triple covering of the largest Fischer group Fi$_{24}'$
as an automorphism group.

It is known that the three-state Potts model at the critical point
can be described as a ``subset'' of the minimal model $\CM(6,5)$ with
$c=\frac{4}{5}$ containing ten primary fields $\phi_{(r,s)}$ of conformal weights
\begin{align}
  h_{r,s}=\frac{(6r - 5s)^2 - 1}{120},
\end{align}
where $1\leq r<5,1\leq s < 6$ and $(r,s)\simeq (5-r,6-s)$.
The $c=\frac{4}{5}$ Virasoro characters labelled by two integers $(r,s)$
are given by
\begin{align}
\label{TP Ch}
  f^{(5)}_{r,s}(q) = \frac{1}{\eta(q)}
  \sum_{n \in \mathbb{Z}} \Big[ q^{\frac{(60n+6r-5s)^2}{120}}
  - q^{\frac{(60n+6r+5s)^2}{120}}\Big].
\end{align}
Some but not all of these primary fields
are present in the critical three-state Potts model.
One can indeed show that a subset of
the ten primary fields are closes under the fusion rules, which leads
to a non-diagonal modular invariant partition function
\begin{align}
  Z = \sum_{r=1,2}  \Big| f^{(5)}_{r,1} +  f^{(5)}_{r,5} \Big|^2
  + 2\Big|  f^{(5)}_{r,3} \Big|^2.
  \label{Potts}
\end{align}
(\ref{Potts}) is the partition function of the three-state Potts model,
which implies that only $\phi_{r,s}$ and two copies of $\phi_{r,3}$ with
$r=1,2$ and $s=1,5$ are present in the theory.
Notice also that the three-state Potts model has $\mathbb{Z}_3$ symmetry
under which two copies of $\phi_{r,3}$ for each $r$ transform differently.
This $\mathbb{Z}_3$ symmetry plays a key role to have well-defined 
fusion rules of the three-states Potts model\cite{DiFrancesco:639405}.

It is natural from (\ref{Potts}) to define the characters of the critical
three-state Potts model as follows,
\begin{align}
\label{PottsCh}
\begin{split}
  f_0(\t) & = f^{(5)}_{1,1}(\t) + f^{(5)}_{1,5}(\t),
  \\
  f_1(\t) & = f^{(5)}_{2,1}(\t) + f^{(5)}_{2,5}(\t),
  \\
  f_2(\t) & = f_2'(\t) = f^{(5)}_{1,3}(\t),
  \\
  f_3(\t) & = f_3'(\t) = f^{(5)}_{2,3}(\t).
\end{split}
\end{align}
The modular S-matrix of the model then becomes
\begin{align}
\label{Smatrix}
\begin{pmatrix}
    f_{0}(-1/\t) \\ f_{1}(-1/\t) \\
    f_{2}(-1/\t) \\ f_2'(-1/\t)  \\
    f_{3}(-1/\t) \\ f_3'(-1/\t)
\end{pmatrix} =
\frac{2}{\sqrt{15}}
\begin{pmatrix}
    -s_1 & s_2 & -s_1 & -s_1 & s_2 & s_2 \\
    s_2 & s_1 & s_2 & s_2 & s_1 & s_1 \\
    -s_1 & s_2 & - \omega s_1 & - \omega^2 s_1 & \omega s_2 & \omega^2 s_2 \\
    -s_1 & s_2 & - \omega^2 s_1 & -\omega s_1 & \omega^2 s_2 & \omega s_2 \\
    s_2 & s_1 & \omega s_2 & \omega^2 s_2 & \omega s_1 & \omega^2 s_1\\
    s_2 & s_1 & \omega^2 s_2 & \omega s_2 & \omega^2 s_1 & \omega s_1
\end{pmatrix}
\begin{pmatrix}
    f_{0}(\t) \\ f_{1}(\t) \\
    f_{2}(\t) \\ f_{2}'(\t) \\
    f_{3}(\t) \\ f_{3}'(\t)
\end{pmatrix}.
\end{align}
where $s_1=\mbox{sin}\left(\frac{6\pi}{5}\right), s_2=\mbox{sin}\left(\frac{12\pi}{5}\right)$ and $\omega = e^{\frac{2 \pi i}{3}}$.

One can show that the characters (\ref{PottsCh}) satisfy
a bilinear relation
\begin{align}
\begin{split}
\label{bilinear Fi24}
    j(\t)-744 &= f_0(\t) \tilde f_0(\t) + f_1(\t)  \tilde f_1(\t) + f_2(\t) \tilde f_2(\t)
    + f'_2(\t) \tilde f'_2(\t) \\
    & +  f_3(\t)  \tilde f_3(\t) +  f'_3(\t) \tilde f'_3(\t),
\end{split}
\end{align}
where the four characters of a dual theory $\tilde{f}_i(\tau)$ are the solutions to a fourth order 
differential equation,
\begin{align}
\label{4th MDE}
    \Big[ \CD_\tau^4  + \mu_1 E_4(\tau) \CD_\tau^2 + \mu_2 E_6(\tau) \CD_\tau  + \mu_3 E_4^2(\tau)\Big]
    \tilde f_i(\tau) = 0.
\end{align}
Here, we denote by $\tilde{f}_{1}(\t), \tilde{f}_{2}(\t), \tilde{f}_{3}(\t)$ and $\tilde{f}_{4}(\t)$ characters of 
weights $h=0$, $\frac{8}{5}$, $\frac{4}{3}$ and $\frac{29}{15}$, respectively. We further use the $q$-expansion of $\tilde f_i(\tau)$
\begin{align}
\label{q expansion}
\tilde f_i(\t) = q^{\left( h_i - \frac{c}{24}\right)} \left( a_0  + a_1 q + a_2 q^2 + a_3 q^3 + \cdots \right),
\end{align}
to fix free parameters $\mu_i$ in \eqref{4th MDE}. The fourth order 
differential equation of our interest then becomes
\begin{align}
\label{4rd MDE dual}
    0 = \Big[ \CD_\tau^4  + \frac{907 \pi^2}{225} E_4(\tau) \CD_\tau^2 -i
    \frac{4289 \pi^3}{675} E_6(\tau) \CD_\tau  -\frac{175769 \pi^4}{50625} E_4^2(\tau)\Big]
    \tilde f(\tau).
\end{align}
Now, It is easy to see that the solutions of \eqref{4rd MDE dual} have the $q$-expansion
\begin{align}
\label{FCCh2}
\begin{split}
    \tilde f_0(\t) &= q^{-\frac{29}{30}}
    \Big(1 + 57478 q^2 + 5477520 q^3 + 201424111 q^4  + \cdots \Big), \\
    \tilde f_1(\t) &=  q^{\frac{19}{30}} \Big(8671 + 1675504 q + 83293626 q^2 + 2175548448 q^3
    + \cdots \Big), \\
    \tilde f_2(\t) & = \tilde f'_2(\t)= q^{\frac{11}{30}} \Big(783 + 306936 q + 19648602 q^2
    + \cdots \Big), \\
    \tilde f_3(\t) &= \tilde f_3'(\t) =  q^{\frac{29}{30}} \Big(64584 + 6789393 q + 261202536 q^2
    + \cdots \Big).
\end{split}
\end{align}

From the bilinear relation (\ref{bilinear Fi24}), one can read that the modular
S-matrix of the dual theory is
\begin{align}
\label{Smatrix}
\begin{pmatrix}
    {\tilde f}_{0}(-1/\t) \\ {\tilde f}_{1}(-1/\t) \\
    {\tilde f}_{2}(-1/\t) \\ {\tilde f}_2'(-1/\t)  \\
    {\tilde f}_{3}(-1/\t) \\ {\tilde f}_3'(-1/\t)
\end{pmatrix} =
\frac{2}{\sqrt{15}}
\begin{pmatrix}
    -s_1 & s_2 & -s_1 & -s_1 & s_2 & s_2 \\
    s_2 & s_1 & s_2 & s_2 & s_1 & s_1 \\
    -s_1 & s_2 & - \omega^2 s_1 & - \omega s_1 & \omega^2 s_2 & \omega s_2 \\
    -s_1 & s_2 & - \omega s_1 & -\omega^2 s_1 & \omega s_2 & \omega^2 s_2 \\
    s_2 & s_1 & \omega^2 s_2 & \omega s_2 & \omega^2 s_1 & \omega s_1\\
    s_2 & s_1 & \omega s_2 & \omega^2 s_2 & \omega s_1 & \omega^2 s_1
\end{pmatrix}
\begin{pmatrix}
    {\tilde f}_{0}(\t) \\ {\tilde f}_{1}(\t) \\
    {\tilde f}_{2}(\t) \\ {\tilde f}_{2}'(\t) \\
    {\tilde f}_{3}(\t) \\ {\tilde f}_{3}'(\t)
\end{pmatrix}.
\end{align}
We can also verify that
the dual theory with $3 \cdot$Fi$_{24}'$ symmetry
has positive integer fusion coefficients,
obtained from the above S-matrix via the Verlinde formula.

Notice that each coefficient of (\ref{FCCh2}) can be expressed as
a sum of dimensions of representations of $3 \cdot$Fi$_{24}'$,
the triple covering of the largest Fischer group. $3 \cdot$Fi$_{24}'$
is one of the maximal subgroup of the Monster group, and is of order $2^{21} \cdot 3^{17} \cdot 5^2 \cdot 7^3 \cdot 11 \cdot 13 \cdot 17 \cdot 23 \cdot 29$.
It has $256$ irreducible representations including
${\bf 783}$, ${\bf 8671}$ and ${\bf 64584}$ that agree with
the first coefficient of $\tilde f_2(\t)$, $\tilde f_1(\t)$ and $\tilde f_3(\t)$.
One can also show that
\begin{align}
\label{FschEx01}
\begin{split}
  57478 & = {\bf 1} \oplus {\bf 57477}, \quad 1675504 = {\bf 8671} \oplus {\bf 1666833},
  \\
  306936 & = {\bf 783} \oplus {\bf 306153}, \quad 6789393 = {\bf 64584} \oplus {\bf 6724809} ,
  \\
  5477520 &= \bf{1} \oplus \bf{57477} \oplus \bf{555611} \oplus \bf{4864431}
\end{split}
\end{align}
where the first term of each line in (\ref{FschEx01}) can
be understood as the Virasoro descendant of the corresponding
primary field.

Now we will find the twined characters of $c=\frac{116}{5}$ CFT and examine 
if they form a bilinear relation analogous to \eqref{extended bilinear}. For instance, the twined characters for $g=2A$ are given by
\begin{align}
\label{2A twined}
\begin{split}
\tilde{f}_0^{2A}(q) &= q^{-\frac{29}{30}} \left( 1 + 1158 q^2 + 20112 q^3  + \cdots \right), \\
\tilde{f}_{1}^{2A}(q) &= q^{\frac{19}{30}} \left( 351 + 11504 q  + \cdots \right) , \\ 
\tilde{f}_{2}^{2A}(q) &= \tilde{f}{'}_{2}^{2A}(q) = q^{\frac{11}{30}} \left( 79 + 2808 q  + \cdots \right), \\
\tilde{f}_{3}^{2A}(q) &= \tilde{f}{'}_{3}^{2A}(q) = q^{\frac{29}{30}} \left( 1352 + 27729 q  + \cdots \right). \\
\end{split}
\end{align}
Combined with the characters of three-state Potts model, it turns out that the twined characters \eqref{2A twined} merged into the Mckay-Thompson series of class 2A.
\begin{align}
\begin{split}
\label{Bilinear relation j2A}
  j^{2A}(\tau) &= f_0(\tau) \cdot \tilde f^{2A}_0(\t) + f_1(\t) \cdot \tilde f^{2A}_1(\tau)  + f_2(\t) \cdot \tilde f^{2A}_2(\tau)  + f{'}_2(\t) \cdot \tilde f{'}^{2A}_2(\tau) \\
                     & \quad  +  f_3(\t) \cdot \tilde f^{2A}_3(\tau) +  f{'}_3(\t) \cdot \tilde f{'}^{2A}_3(\tau)
\end{split}
\end{align}
As another example, twined characters for $g = 3E$ read
\begin{align}
\label{3E twined}
\begin{split}
\tilde{f}_0^{3E}(q) &= q^{-\frac{29}{30}} \left( 1 + 616 q^2 + 7833 q^3  + \cdots \right), \\
\tilde{f}_{1}^{3E}(q) &= q^{\frac{19}{30}} \left( -77 - 1925 q  + \cdots \right) , \\ 
\tilde{f}_{2}^{3E}(q) &= q^{\frac{11}{30}} \left( 54 \alpha + 1485 \alpha q  + \cdots \right), \\
\tilde{f}{'}_{2}^{3E}(q) &= q^{\frac{11}{30}} \left( 54 \overline{\alpha} + 1485 \overline{\alpha} q  + \cdots \right), \\
\tilde{f}_{3}^{3E}(q) &= q^{\frac{29}{30}} \left( -297 \alpha - 4158 \alpha q  + \cdots \right), \\
\tilde{f}{'}_{3}^{3E}(q) &= q^{\frac{29}{30}} \left( -297 \overline{\alpha} -4158 \overline{\alpha} q  + \cdots \right), \\
\end{split}
\end{align}
where $\alpha = \frac{-1+i\sqrt{3}}{2}$ and $\overline{\alpha}$ is its complex conjugate.
Then, we get a new type of bilinear relation of the form, 
\begin{align}
\begin{split}
  j^{3A}(\tau) &= f_0(\tau) \cdot \tilde f^{3E}_0(\t) + f_1(\t) \cdot \tilde f^{3E}_1(\tau) + f_2(\t) \cdot \tilde f^{3E}_2(\tau)  + f'_{2}(\t) \cdot \tilde f{'}^{3E}_{2}(\tau) \\
               &\quad  + f_3(\t) \cdot \tilde f^{3E}_3(\tau)  +  f'_{3}(\t) \cdot \tilde f{'}^{3E}_{3}(\tau)
\end{split}
\end{align}
where $j^{3A}(\tau)$ is Mckay-Thompson series of class $3A$ given by,
\begin{align}
\begin{split}
j^{2A}(\tau) &= \left( \left( \frac{\eta(q)}{\eta(q^3)} \right)^{12} + 27 \right)^2 \Big/ \left( \frac{\eta(q)}{\eta(q^3)} \right)^{12} -42 \\
                   &= \frac{1}{q} +783 q + 8672 q^2 + 65467 q^3 + 371520 q^4 + \cdots.
\end{split}
\end{align}

In appendix \ref{App:C}, we listed generalized bilinear relations that 
various twined characters satisfy. These suggest that the six-character
rational CFT with $c=\frac{116}{5}$ dual to the critical three-state
Potts model has $3 \cdot$Fi$_{24}'$ as an automorphism group.

\section{Dual of the Critical Ising$^2$ and $2 \cdot 2^{1+22} \cdot \text{Co}_{2}$ } \label{Sec : 4}
We now in turn consider a rational CFT dual to the tensor product of the
critical Ising model, the simplest example of the $c=1$
CFTs studied in \cite{Elitzur1987,Ginsparg1988b, Dijkgraaf1989}.  This theory has nine primaries of conformal weights
\begin{align}
h_1 = h'_1 = \frac{1}{2}, \ h_2 = h'_2 =\frac{1}{16}, \ h_3 = 1, \ h_4 = h'_4 = \frac{9}{16}, \ h_5 = \frac{1}{8},
\end{align}
and their characters can be expressed by characters of the critical Ising model \eqref{Ising Ch}.
\begin{align}
\label{c=1 Ch}
\begin{split}
&g_0(\tau) = f_0(\tau) \cdot f_0(\tau), \ \ g_1(\tau) = g'_1(\tau) =f_{0}(\tau) \cdot f_{\epsilon}(\tau), \\
&g_2(\tau) = g'_2(\tau) = f_{0}(\tau) \cdot f_{\sigma}(\tau), \ \ g_3(\tau) = f_{\epsilon}(\tau) \cdot f_{\epsilon}(\tau), \\
&g_4(\tau) = g'_4(\tau) = f_{\sigma}(\tau) \cdot f_{\epsilon}(\tau), \ \ g_5(\tau) = f_{\sigma}(\tau) \cdot f_{\sigma}(\tau)
\end{split}
\end{align}
The $9 \times 9$ extended modular matrix $\mathbb{S}$ reads,
\begin{align}
\label{Double Ising S-law}
\mathbb{S} =
 \frac{1}{4}
\begin{pmatrix}
1 & 1 & 1 & \sqrt{2} & \sqrt{2} & 1 & \sqrt{2} & \sqrt{2} & 2\\
1 & 1 & 1 & \sqrt{2} & -\sqrt{2} &1  & \sqrt{2}& -\sqrt{2}& -2\\
1 & 1 & 1 & -\sqrt{2} & \sqrt{2} &1  & -\sqrt{2}& \sqrt{2}& -2\\
\sqrt{2} & \sqrt{2}& -\sqrt{2} & 2 & 0 & -\sqrt{2} & -2 & 0 & 0\\
\sqrt{2} & -\sqrt{2} & \sqrt{2} & 0 & 2 & -\sqrt{2} & 0& -2& 0\\
1 & 1 & 1 & -\sqrt{2} & -\sqrt{2} & 1& -\sqrt{2}& -\sqrt{2}& 2 \\
\sqrt{2} & \sqrt{2} & -\sqrt{2} & -2 & 0 & -\sqrt{2} & 2& 0& 0\\
\sqrt{2} & -\sqrt{2} & \sqrt{2} & 0 & -2 & -\sqrt{2} & 0 & 2& 0\\
2 & -2 & -2 & 0 & 0 & 2 & 0& 0& 0
\end{pmatrix},
\end{align}
thus the product of two critical Ising model admit diagonal modular invariants of the form
\begin{align}
\label{diagonal01}
\begin{split}
Z(\tau, \bar{\tau}) &= |g_0(\tau)|^2 +  |g_1(\tau)|^2 + |g_1'(\tau)|^2 +  |g_2(\tau)|^2 +  |g_2'(\tau)|^2 \\
                                & \quad + |g_3(\tau)|^2 + |g_4(\tau)|^2 + |g_4'(\tau)|^2 + |g_5(\tau)|^2.
\end{split}
\end{align}
We assumed that the characters \eqref{c=1 Ch} satisfy a bilinear relation
\begin{align}
\begin{split}
\label{Bilinear Ising sq}
j(\tau) - 744 &= g_0(\tau)  \tilde{g}_0(\tau) + g_1(\tau)  \tilde{g}_1(\tau)  + g_1'(\tau)  \tilde{g}_1'(\tau) + g_2(\tau)  \tilde{g}_2(\tau) + g_2'(\tau)  \tilde{g}_2'(\tau)  \\
& \quad +  g_3(\tau)  \tilde{g}_3(\tau) + g_4(\tau)  \tilde{g}_4(\tau) + g_4'(\tau)  \tilde{g}_4'(\tau) + g_5(\tau)  \tilde{g}_5(\tau)
\end{split}
\end{align}
where the characters of a dual CFT with weights
\begin{align}
\label{c23 weights}
\tilde{h}_1 = \tilde{h}_1' = \frac{3}{2}, \ \tilde{h}_2 = \tilde{h}_2' =\frac{31}{16}, \ \tilde{h}_3 = 1, \ \tilde{h}_4 = \tilde{h}_4' = \frac{23}{16}, \ \tilde{h}_5 = \frac{15}{8}.
\end{align}
are the solutions of a differential equation,
\begin{align}
\begin{split}
\label{6thMDE}
  0 & = \Big[ \CD_\tau^6  + \mu_1 E_4(\tau) \CD_\tau^4 + \mu_2 E_6(\tau) \CD_\tau^3  + \mu_3 E_4^2(\tau) \CD_\tau^2
  + \mu_4 E_4(\tau) E_6(\tau) \CD_\tau
  \\ &
  + \mu_5 E_4^3(\tau) + \mu_6 E^2_6(\tau) +\mu_7 \frac{E_4^2(\tau)}{E_6(\tau)} \CD_\tau^5
  + \mu_8  \frac{E_4^3(\tau)}{E_6(\tau)}\CD_\tau^3 + \mu_9  \frac{E_4^4(\tau)}{E_6(\tau)} \CD_\tau \Big] \tilde{g}(\tau).
\end{split}
\end{align}
However, inserting $q$-expansion of the characters into \eqref{6thMDE} does not determine all $\mu_i$ in \eqref{6thMDE}, because 
it give us six constraints while there are nine unfixed parameters in \eqref{6thMDE}. 
To remedy it, we compared two bilinear relations \eqref{Bilinear Ising sq} and \eqref{Ising BM}, which eventually provide us three additional constraints
\begin{align}
\begin{split}
\label{3 additional}
\tilde{f}_0(\t) &= f_0(\tau) \tilde{g}_0(\tau) + f_{\epsilon}(\tau) \tilde{g}_1(\tau) + f_{\sigma}(\tau) \tilde{g}_2(\tau)\\
\tilde{f}_{\epsilon}(\t) &= f_0(\tau)  \tilde{g}_1'(\tau) + f_{\epsilon}(\tau) \tilde{g}_3(\tau) + f_{\sigma}(\tau) \tilde{g}_4(\tau) \\
\tilde{f}_{\sigma}(\t) &= f_0(\tau) \tilde{g}_2'(\tau) + f_{\sigma}(\tau) \tilde{g}_4'(\tau) + f_{\epsilon}(\tau) \tilde{g}_5(\tau).
\end{split}
\end{align}
Now one can fix nine parameters $\mu_i$ combining three equations \eqref{3 additional} with six constraints from \eqref{6thMDE}. 
In this way, we find that the  nine parameters $\mu_i$ read,
\begin{align}
\begin{split}
\mu_1 &= \frac{2647 \pi ^2}{576}, \ \mu_2 = -i\frac{3495263687743883 \pi ^3}{140037228850176}, \ \mu_3 = -\frac{598979 \pi ^4}{82944}, \\
\mu_4 &= -i\frac{779163580240684865 \pi ^5}{20165360954425344}, \ \mu_5 = \frac{634818358457751325 \pi ^6}{13443573969616896},  \\
\mu_6 &= -\frac{5641332583789180993 \pi ^6}{120992165726552064}, \  \mu_7 = i \pi, \ \mu_8 = i \frac{810490346954549 \pi ^3}{46679076283392},\\
\mu_9 &= i \frac{284686915225948007 \pi ^5}{6721786984808448}.
\end{split}
\end{align}
As a result, we finally get below six different characters of dual CFT.
\begin{align}
\label{DDI ch}
\begin{split}
\tilde{g}_0(\tau) &= q^{-\frac{23}{24}}\left(1 + 46851 q^2 + 4310154 q^3 +
 155027130 q^4  + \cdots \right)\\
\tilde{g}_1(\tau) &= \tilde{g}'_1(\tau)= q^{\frac{13}{24}} \left(2300  + 529828 q + 28051444 q^2  +\cdots \right)\\
\tilde{g}_2(\tau) &= \tilde{g}'_2(\tau) = q^{\frac{47}{48}} \left(47104  + 4757504 q + 178382848 q^2  + \cdots \right) \\
\tilde{g}_3(\tau) &= q^{\frac{1}{24}} \left(23 + 46598 q + 4311948 q^2 +
 155017746 q^3  + \cdots \right) \\
\tilde{g}_4(\tau) &= \tilde{g}'_4(\tau) = q^{\frac{23}{48}} \left(2048 + 565248 q + 31700992 q^2  + \cdots \right) \\
\tilde{g}_5(\tau) &= q^{\frac{11}{12}} \left(47104 + 5230592 q + 204855296 q^2 +
 4630417408 q^3  + \cdots \right)
\end{split}
\end{align}

One can easily show that every coefficient in \eqref{DDI ch}
can be expressed as a sum of the dimension of the irreducible representation of
$2 \cdot 2^{1+22} \cdot $Co$_2$. 
$2 \cdot 2^{1+22} \cdot $Co$_2$ is a maximal subgroup of the baby Monster group.
Its order is $2^{15}\cdot3^6\cdot5^3\cdot 7\cdot 11\cdot 23$. This multi-covering of the second largest
Conway group $\text{Co}_2$ has irreducible representations that includes ${\bf 23}$, ${\bf 2048}$, ${\bf 2300}$ and ${\bf 47104}$. These numbers are in perfect agreement with
the first terms of $\tilde g_3(\t)$, $\tilde g_4(\t)$, $\tilde g_1(\t)$, $\tilde g_2(\t)$ and $\tilde g_5(\t)$.
Referring the table \ref{degrees of irreps}, the other numbers in the characters can be decomposed as follow,
\begin{align}
\label{CochEx01}
\begin{split}
  46851 & = {\bf 1} \oplus {\bf 275} \oplus {\bf 46575}, \quad
  529828 = {\bf 2300} \oplus {\bf 50600} \oplus {\bf 476928},
  \\
  4757504 & = 2 \cdot {\bf 47104} \oplus {\bf 4663296}, \quad  46598 = {\bf 23} \oplus {\bf 46575}, \\
  565248  &= {\bf 2048} \oplus {\bf 563200},\\
 5230592 &=   {\bf{2048}} \oplus {\bf{47104}} \oplus {\bf{518144}} \oplus  {\bf{4663296}} \\
4310154 &=  {\bf{1}} \oplus  {\bf{253}} \oplus {\bf{275}} \oplus {\bf{46575}} \oplus {\bf{1024650}} \oplus {\bf{3238400}},
\end{split}
\end{align}
where the first term in each decomposition can be again understood as the Virasoro descendent
of the corresponding primary state. Thus we conjecture that the nine-character rational CFT of $c=23$, dual to the product of two critical Ising model, has $2 \cdot 2^{1+22} \cdot \text{Co}_2$ as an automorphism group.

The modular S-matrix of the $c=23$ CFT is identical to the \eqref{Double Ising S-law}, because of the bilinear relation \eqref{Bilinear Ising sq}. Therefore, the modular invariant partition function of $c=23$ CFT is given by
\begin{align}
\label{Conway partition}
\begin{split}
Z(\tau, \bar{\tau}) &= |\tilde{g}_0(\tau)|^2 +  |\tilde{g}_1(\tau)|^2  +  |\tilde{g}'_1(\tau)|^2 +  |\tilde{g}_2(\tau)|^2  +  |\tilde{g}'_2(\tau)|^2  \\
                                & \quad + |\tilde{g}_3(\tau)|^2 + |\tilde{g}_4(\tau)|^2 + |\tilde{g}'_4(\tau)|^2 + |\tilde{g}_5(\tau)|^2.
\end{split}
\end{align}
Also, the modular S-matrix \eqref{Double Ising S-law} guarantees the positive integer fusion rule algebra coefficients.

We also propose that the characters of the above RCFT \eqref{DDI ch} obey
intriguing bilinear relations with those of the critical
Ising model \eqref{Ising Ch} to give the baby Monster modules \eqref{BM characters},
\begin{align}
\begin{split}
  \tilde f_0(\t) & = f_0(\t) \tilde g_0(\t) + f_\e(\t) \tilde g_1(\t) + f_\s(\t) \tilde g_2(\t),
  \\
  \tilde f_\e(\t) & = f_0(\t) \tilde g_1(\t) + f_\e(\t) \tilde g_3(\t) +  f_\s(\t) \tilde g_4(\t),
  \\
  \tilde f_\s(\t) & = f_0(\t) \tilde g_2(\t) + f_\e(\t) \tilde g_4(\t) + f_\s(\t) \tilde g_5(\t).
\end{split}
\end{align}
%

\section{Self-Dual RCFT and $2\cdot\text{Co}_1$} 

We discuss in this section a ``self-dual'' RCFT with
$c=12$ whose torus partition function admits a
natural decomposition in terms of dimensions of
representation of the Conway group $2 \cdot$Co$_1$.

The RCFT of our interest is self-dual in a sense that its three
characters, denoted by $f_0(\t), f_1(\t)$ and $f_2(\t)$,
satisfy a bilinear relation giving, 
\footnote{Strictly speaking, a bilinear relation \eqref{BR for self-dual RCFT} 
cannot be an example of Monster anatomy, because $j(\tau) + 96$ is not the Monster module. 
Nonetheless, in this section, we discuss a self-dual RCFT of $c=12$ because partition function of this theory 
also exhibit moonshine for Conway group.
}
\begin{align}
\label{BR for self-dual RCFT}
     j(\tau) + 96 = f_0^2(\t) + f_1^2(\t) + \frac12 f_2^2(\t)
%
%
%
\end{align}
with $\l(\t) = \frac{\theta_{2}^4(\tau)}{\theta_{3}^4(\tau)}$ and
\begin{align}
\label{c=12 solutions}
\begin{split}
  f_0(\t) & = \Big( -\frac{1}{2}\l^3(\t) + \frac{3}{2}\l^2(\t) - \frac{3}{2}\l(\t)  + 2 \Big)
  \Big(\frac{16}{\l(\t)\big(1-\l(\t)\big)}\Big),
  \\ & =
  q^{-1/2} \Big( 1 + 276 q + 11202 q^2 + 184024q^3 + \cdots \Big),
  \\
  f_1(\t) & =   \frac{8 \l^3(\t)}{\l(\t)\big(1-\l(\t)\big)},
  \\ & =
  2048 q \Big( 1 + 24 q + 300 q^2 + 2624 q^3 + \cdots \Big),
  \\
  f_2(\t) & =
  \frac{16\l^3(\t) }{\l(\t)\big(1-\l(\t)\big)} + 24 ,
  \\ &=
  24 + 4096q + 98304q^2 + 1228800 q^3 + \cdots.
\end{split}
\end{align}
Here $f_0(\t)$ is the vacuum character while $f_1(\t)$ and $f_2(\t)$
are characters for primary states of conformal weight $h=\frac{3}{2}$ and $h=\frac{1}{2}$.
These characters are three independent solutions  to a modular differential equation below,
\begin{align}
\label{c=12 MDE}
\begin{split}
 \left[ \partial_\tau^3 - \frac{1}{2} E_2(\tau) \partial_\tau^2 + \frac{1}{24} \Big(
 E_2^2(\tau) - 13 E_4(\tau) \Big)
  \partial_\tau \right] f(\tau) = 0.
\end{split}
\end{align}
For later convenience, let us define
$f_{--}(\t)$, $f_{-+}(\t)$ and $f_{+-}(\t)$ as follows
\begin{align}
\label{c=12 sols}
\begin{split}
    f_{--}(\t) & = f_0(\t) + f_1(\t) ,
    \\ & =
    q^{-1/2} \Big( 1 + 276 q + 2048q^{3/2} + 11202q^2 + 49152 q^{5/2} + \cdots \Big),
    \\
    f_{-+}(\t) & = f_0(\t) - f_1(\t),
    \\ & =
    q^{-1/2} \Big( 1 + 276 q - 2048q^{3/2} + 11202q^2 - 49152 q^{5/2} + \cdots \Big),
    \\
    f_{+-}(\t) & = f_2(\t)
    \\ & =
    24 + 4096 q + 98304q^2 + 1228800q^3 + 10747904q^4+ \cdots.
\end{split}
\end{align}

From the modular S-matrix of the self-dual theory,
\begin{align}
\label{S-matrix c12}
\begin{pmatrix}
    f_{--}(1-\lambda) \\ f_{-+}(1-\lambda) \\ f_{+-}(1-\lambda)
\end{pmatrix}
=
\begin{pmatrix}
    1 & 0& 0\\  0 & 0& 1 \\ 0 & 1 & 0
\end{pmatrix}
\begin{pmatrix}
    f_{--}(\lambda) \\ f_{-+}(\lambda) \\ f_{+-}(\lambda)
\end{pmatrix}.
\end{align}
one can show that the $SL(2,\mathbb{Z})$ invariant partition function would be
\begin{align}
  Z = \frac12 \Big( \big| f_{--}(\t) \big|^2 + \big| f_{-+}(\t) \big|^2 + \big| f_{+-}(\t) \big|^2 \Big)
  + \text{const.}
  \label{PartitionConway1}
\end{align}

Notice here that the character $f_{--}(\t)$ is nothing but
the Neveu-Schwarz (NS) partition function $K(\t)$ of $\CN=1$ extremal superconformal
theory\cite{Witten2007}. In fact, one can understand from (\ref{PartitionConway1}) that
the RCFT with $c=12$ of our interest is the GSO projection of the
$\CN=1$ extremal SCFT where
\begin{align}
\begin{split}
  f_{--}(\t) & = \text{tr}_{\text{NS}} \Big[ q^{L_0- c/24} \Big],
  \\
  f_{-+}(\t) & = \text{tr}_{\text{NS}} \Big[ (-1)^F q^{L_0- c/24} \Big],
  \\
  f_{+-}(\t) & = \text{tr}_{\text{R}} \Big[ q^{L_0- c/24} \Big],
\end{split}
\end{align}
and the constant term corresponding to the Witten index, $\text{tr}_{\text{R}}
\big[(-1)^F q^{L_0 - c/24} {\bar q}^{{\bar L}_0 - c/24}\big]= 24^2$.
Here the each trace can be performed in either the NS Hilbert space
or the Ramond Hilbert space of the $\CN=1$ SCFT.
This SCFT was first made by Frenkel, Lepowsky and Meurman \cite{Frenkel1988},
and revisited later by Duncan \cite{Duncan2005,Duncan2014a}.

The $\CN=1$ extremal SCFT with $c=12$
is well-known to have  $2\cdot$Co$_{1}=$Co$_0$ as an automorphism group.
After the GSO projection, the double covering
of the largest Conway group continues to serve as an automorphism group of
the self-dual theory.
The Virasoro character decomposition of the
partition further gives new evidence that the automorphism group of
the RCFT with $c=12$ may be enhanced to a larger group $2^{1+24}\cdot\text{Co}_1$,
a maximal subgroup of the Monster group.
To see this, let us decompose the partition
function (\ref{PartitionConway1}) in terms of the $c=12$ Virasoro characters,
\begin{align}
\label{c=12 partition}
\begin{split}
    Z(\t,\bar{\t}) & =
    \chi_0(q) \bar{\chi}_0(q)
    + 276 \Big( \chi_{0}(\bar{q}) \bar{\chi}_{1}(\bar{q}) +  \text{c.c.} \Big)
    + 76176 \chi_{1}(\bar{q}) \bar{\chi}_{1}(\bar{q})
    \\ &
    + 10925 \Big( \chi_{0}(\bar{q}) \bar{\chi}_{2}(\bar{q}) + \text{c.c.} \Big)
    + 3015300 \Big( \chi_{1}(\bar{q}) \bar{\chi}_{2}(\bar{q}) + \text{c.c.} \Big)
    \\ &
    + 1081344 \Big(\chi_{\frac{1}{2}}(q) \bar{\chi}_{\frac{5}{2}}(\bar{q}) + \text{c.c.} \Big)
    + 49152 \chi_{\frac{1}{2}}(\bar{q}) \bar{\chi}_{\frac{1}{2}}(\bar{q}) + \cdots.
\end{split}
\end{align}
It turns out that every coefficients in \eqref{c=12 partition} is related to the
irreducible representations of $2^{1+24} \cdot$ Co$_{1}$.
Some details of the number decomposition are presented below.
\begin{align}
\begin{split}
&49152= {\bf{1}} \oplus 2 \cdot {\bf{276}} \oplus {\bf{299}} \oplus {\bf{1771}} \oplus {\bf{8855}} \oplus {\bf{37674}},  \\
&276 = {\bf{23}} \oplus {\bf{253}}, \quad 10925 = {\bf{299}} \oplus {\bf{1771}} \oplus {\bf{8855}} \oplus {\bf{10626}}, \\
&76176 =  {\bf{1}} \oplus {\bf{276}} \oplus {\bf{299}} \oplus {\bf{1771}} \oplus {\bf{8855}} \oplus {\bf{27300}} \oplus {\bf{37674}}
\end{split}
\end{align}

\section{Discussion}

It is known that the CFT of $c=1$ 
discussed in section \ref{Sec : 4} allows 
a non-diagonal partition function 
\begin{align}
\label{nd partition function}
\begin{split}
Z(\tau, \bar{\tau}) = |g_0(\tau) + g_3(\tau)|^2 + |g_1(\tau) + g'_1(\tau)|^2 + 2 |g_5(\tau)|^2,
\end{split}
\end{align}
different from (\ref{diagonal01}). 
(\ref{nd partition function}) is the partition 
function of another example of $c=1$ CFT studied 
in \cite{Ginsparg1988b,Dijkgraaf1989}. One can show that the characters 
in (\ref{nd partition function}) can obey 
a new bilinear relation
\begin{align}
\label{bilienar nd ising}
\begin{split}
 j(\tau) - 720 &= 
 \big(g_0(\tau) + g_3(\tau)\big) \big(\tilde{g}_0(\tau) + \tilde{g}_3(\tau)\big) 
 + \big(g_1(\tau) + g'_1(\tau)\big) \big(\tilde{g}_1(\tau) + \tilde{g}'_1(\tau)\big) 
 \\ & + 2 g_5(\tau) \tilde{g}_5(\tau), 
\end{split}
\end{align}
from which one can read the partition function of the dual 
theory 
\begin{align}
\label{aaa}
  \tilde Z(\tau, \bar{\tau}) =
  |\tilde g_0(\tau) + \tilde g_3(\tau)|^2 + |\tilde g_1(\tau) + \tilde g'_1(\tau)|^2
  + 2 |\tilde g_5(\tau)|^2. 
\end{align}
Notice that (\ref{aaa}) can be understood as a non-diagonal 
partition function of the rational CFT of $c=23$
discussed in section \ref{Sec : 4}. It is obvious 
that the dual CFT also exhibit the moonshine for $2 \cdot 2^{1+22} \cdot $Co$_2$. 
This is somewhat trivial, because it shares the characters \eqref{DDI ch} as building blocks.
However, above example cannot be considered as a class of the Monster anatomy,
because $j(\tau)-720$ is not the Monster module.

It has been proposed that the Hecke images of the vector-valued modular form can construct a set of admissible characters for specific RCFT \cite{Harvey2018}. The Hecke image $T_p f_i(\tau)$ is characterized by the conductor $N$ and a natural number $p$ that is relatively prime to $N$.  
As an example, it has been known that the conductor of the critical Ising model is given by $N = 48$. Then, one can show that the Hecke images for $p=47$ are exactly agree to the characters of the baby Monster CFT, namely \eqref{BM characters}. In similar way, it is easy to see that the conductor of three-state Potts model and the tensor product of the critical Ising model are given by $N=30$ and $N=24$, respectively. We checked that the characters \eqref{FCCh2} can be considered as the Hecke images of \eqref{PottsCh} with $p=29$. In similar way, we found that the vector-valued modular form $\left( \tilde g_0(\tau) + \tilde g_3(\tau) , \tilde g_1(\tau) + \tilde g'_1(\tau), \tilde g_5(\tau) \right)$ in \eqref{aaa} is also realized as the Hecke images of  $\left(  g_0(\tau) +  g_3(\tau) ,  g_1(\tau) +  g'_1(\tau),  g_5(\tau) \right)$ with $p=23$. However, it turns out that the Hecke images of \eqref{c=1 Ch} cannot generate the six characters in \eqref{DDI ch}. More precisely, the characters $\tilde{g_2}(\tau)$ and $\tilde{g_4}(\tau)$ in \eqref{DDI ch} are not able to realized as the Hecke images of the characters in \eqref{c=1 Ch}.

One can also ask if the dual CFT of the tricritical Ising model, the next simplest unitary minimal model $\CM(5,4)$, can exhibit the moonshine phenomena. 
However, it is nontrivial to obtain the characters of the candidate dual CFT with 
$c=\frac{233}{10}$. This is partially because
one cannot determine all free parameters in 
the corresponding MDE completely from the known CFT data\cite{Hampapura2016}. 
We tried to find the dual characters of $c=\frac{233}{10}$ CFT using the Hecke operator of $N=240$ and $p=233$, however it turns out that this Hecke operator does not produce admissible characters.

On the other hand, it is known that the tricritical Ising model is endowed with the $\mathcal{N}=1$ supersymmetry. The NS partition function of the model can be contributed by 
the NS superconformal characters of the vacuum and the primary state of 
$h=\frac{1}{10}$. It would be interesting to search for a rational SCFT 
whose NS characters obey a new bilinear relation with those of the tricritical 
Ising model to give the $K(\t)$-function \cite{Witten2007}. This new bilinear 
relation would lead to a picture of the Conway group decomposition, instead of the Monster group decomposition. We leave them as a future project.

The bigger challenge is to find all the two-dimensional dual CFT pairs for 
the rest of the sporadic groups and understand the origin of bilinear relations. In particular, it would be extremely 
interesting to see if a CFT dual to 
`(three-state Potts model)$\times$(a certain CFT)' can show moonshine for the multiple-covering
of the largest Mathieu group, $2^{12}\cdot \text{M}_{24}$. 
This is analogous to the idea used to find the dual CFT with $2\cdot2^{1+22}\cdot \text{Co}_2$ and correspond to the green arrow in figure \ref{Monster Decomposition}.   

\section*{Acknowledgments}
We would like to thank Hyun Kyu Kim for useful discussions. We grateful to the anonymous referee of JHEP for giving us numerous, invaluable comments on the manuscript. KL is supported in part by the National Research Foundation of Korea Grant NRF-2017R1D1A1B06034369.  KL would like to thank the colleagues  at IIP Natal Brazil where the part of work is done. The research of S.L. is supported in part by the National Research Foundation of Korea (NRF) Grant NRF-$2017$R$1$C$1$B$1011440$. K.L. and S.L. thank the Aspen Center for Physics (supported by National Science Foundation grant PHY-1607611).

\newpage
\appendix
\section{Dimension of the irreducible representations} \label{App:A}

Here we give a partial list of the dimension of the irreducible representations for various sporadic groups. 

\begin{table}[h!]
\centering
\renewcommand{\arraystretch}{1.1}
\begin{tabular}{|c|l|}
\hline
  & \multicolumn{1}{|c|}{\rule{0pt}{2.8ex}Dimension of the irreducible representations} \rule[-1.1ex]{0pt}{0pt} \\ \hline \hline
\multirow{ 3}{*}{$2\cdot \mathbb{B}$} & \rule{0pt}{3.8ex}$\{ {\bf{1, 4371, 96255, 96256, 1139374, 9458750, 9550635, 10506240,}}$  \rule{0pt}{3.5ex}\\
& $ {\bf{63532485,347643114, 356054375, 410132480, 1407126890,}}$  \\
& $ {\bf{3214743741,4221380670,4275362520, 4622913750, \cdots}} \}$ \rule[-2.1ex]{0pt}{0pt}  \\ \hline
\multirow{ 3}{*}{$3\cdot \mbox{Fi}_{24}'$} & \rule{0pt}{3.8ex}$\{ {\bf{1, 783, 8671, 57477, 64584, 249458, 306153, 555611, 1603525, }}$  \rule{0pt}{3.5ex}\\
& $ {\bf{1666833, 4864431, 6724809, 19034730, 25356672, 32715683, }}$  \\
& $ {\bf{ 35873145, 40536925, 43779879, 48893768, 74837400, \cdots, }} \}$ \rule[-2.1ex]{0pt}{0pt}  \\ \hline
\multirow{ 4}{*}{$2 \cdot \mbox{Co}_{1}$} & \rule{0pt}{3.8ex}$\{ {\bf{1, 24, 276, 299, 1771, 2024, 2576, 4576, 8855, 17250, 27300, 37674,}}$  \rule{0pt}{3.5ex}\\
& $ {\bf{ 40480, 44275, 80730, 94875, 95680, 170016, 299000, 313950, 315744,}}$  \\
& $ {\bf{345345, 351624, 376740, 388080, 483000, 644644, 673750, 789360,}}$  \\
& $ {\bf{822250, 871884, 1434510, 1450449, 1771000, 1821600, 1841840,\cdots}} \}$ \rule[-2.1ex]{0pt}{0pt} \\ \hline
\multirow{ 4}{*}{$2^{1+24} \cdot \mbox{Co}_{1}$} & \rule{0pt}{3.8ex}$\{ {\bf{1, 276, 299, 1771, 8855, 17250, 27300, 37674, 44275, 80730, 94875,}}$  \rule{0pt}{3.5ex}\\
& $ {\bf{ 98280, 98304, 313950, 345345, 376740, 483000, 644644, 673750,}}$  \\
& $ {\bf{822250, 871884, 1434510, 1450449, 1771000, 1821600, 2055625,}}$  \\
& $ {\bf{ 2260440, 2417415, 2464749, 2464749, 2816856, 2877875, \cdots}} \}$ \rule[-2.1ex]{0pt}{0pt} \\ \hline
\multirow{6}{*}{$2^{1+22} \cdot \mbox{Co}_{2}$} & \rule{0pt}{3.8ex}$\{ {\bf{1, 23, 253, 275, 1771, 2024, 2048, 2277, 2300, 4025, 7084, 9625,  }}$  \rule{0pt}{3.5ex}\\
& $ {\bf{10395, 12650, 23000, 31625, 31878, 37422, 44275, 46575, 47104,  }}$  \\
& $ {\bf{ 50600, 63250, 91125, 113850, 129536, 177100, 184437, 212520,  }}$  \\
& $ {\bf{221375, 226688, 239085, 245916, 253000, 284625, 312984, 368874,}}$  \\
& $ {\bf{ 398475, 430353, 442750, 462000, 467775, 476928, 518144, 531300,}}$  \\
& $ {\bf{ 558900, 563200, 579600, \cdots }} \}$ \rule[-2.1ex]{0pt}{0pt} \\ \hline
\end{tabular}
\caption{Dimension of the irreducible representation in various sporadic groups. The list of the representation read from GAP package \cite{GAP4}.}
\label{degrees of irreps}
\end{table}

\section{Character Tables of $2 \cdot \mathbb{B}$ and $3 \cdot \text{Fi}_{24}'$}\label{App:B} 

\begin{landscape}
\begin{table}[h!]
\resizebox{23.0cm}{!}{
\centering
\renewcommand{\arraystretch}{1.1}
\begin{tabular}{|c||r|r|r|r|r|r|r|r|r|r|r|r|r|r|r|r|}
\hline
 {$n$}  &$1A$ &$2A$ &$2B$ &$2C$ &$2D$ &$4A$ &$3A$ &$6A$ &$3B$ &$6B$ &$4B$ &$4C$ &$4D$ &$8A$ &$5A$ &$10A$\\ \hline \hline
\rule{0pt}{3ex} \rule[-1ex]{0pt}{0pt} {$1$} & $ {\bf{1}}$ & $ {\bf{1}}$  & $ {\bf{1}}$ & $ {\bf{1}}$ & $ {\bf{1}}$ & $ {\bf{1}}$ & $ {\bf{1}}$ & $ {\bf{1}}$ & $ {\bf{1}}$ & $ {\bf{1}}$ & $ {\bf{1}}$ & $ {\bf{1}}$ & $ {\bf{1}}$ & $ {\bf{1}}$ & $ {\bf{1}}$ & $ {\bf{1}}$  \\ \hline
\rule{0pt}{3ex} \rule[-1ex]{0pt}{0pt} {$2$} & $ {\bf{4371}}$ & $ {\bf{4371}}$  & $ {\bf{-493}}$ & $ {\bf{275}}$ & $ {\bf{275}}$ & $ {\bf{-53}}$ & $ {\bf{78}}$ & $ {\bf{78}}$ & $ {\bf{-3}}$ & $ {\bf{-3}}$ & $ {\bf{-77}}$ & $ {\bf{51}}$ & $ {\bf{19}}$ & $ {\bf{-1}}$ & $ {\bf{21}}$ & $ {\bf{21}}$\\ \hline
\rule{0pt}{3ex} \rule[-1ex]{0pt}{0pt} {$3$} & $ {\bf{96255}}$ & $ {\bf{96255}}$   & $ {\bf{4863}}$ & $ {\bf{2047}}$ & $ {\bf{2047}}$ & $ {\bf{103}}$ & $ {\bf{351}}$ & $ {\bf{351}}$ & $ {\bf{27}}$ & $ {\bf{27}}$ & $ {\bf{351}}$ & $ {\bf{223}}$ & $ {\bf{-1}}$ & $ {\bf{-1}}$ & $ {\bf{55}}$ & $ {\bf{55}}$\\ \hline
\rule{0pt}{3ex} \rule[-1ex]{0pt}{0pt} {$4$} & $ {\bf{1139374}}$ & $ {\bf{1139374}}$   & $ {\bf{-26962}}$ & $ {\bf{8878}}$ & $ {\bf{8878}}$ & $ {\bf{782}}$ & $ {\bf{1000}}$ & $ {\bf{1000}}$ & $ {\bf{-53}}$ & $ {\bf{-53}}$ & $ {\bf{-978}}$ & $ {\bf{558}}$ & $ {\bf{-82}}$ & $ {\bf{-26}}$ & $ {\bf{99}}$ & $ {\bf{99}}$\\ \hline
\rule{0pt}{3ex} \rule[-1ex]{0pt}{0pt} {$5$} & $ {\bf{9458750}}$  & $ {\bf{9458750}}$  & $ {\bf{118846}}$ & $ {\bf{37950}}$ & $ {\bf{37950}}$ & $ {\bf{3486}}$ & $ {\bf{2729}}$ & $ {\bf{2729}}$ & $ {\bf{-25}}$ & $ {\bf{-25}}$ & $ {\bf{2750}}$ & $ {\bf{1214}}$ & $ {\bf{318}}$ & $ {\bf{-26}}$ & $ {\bf{175}}$ & $ {\bf{175}}$\\ \hline
\rule{0pt}{3ex} \rule[-1ex]{0pt}{0pt} {$6$} & $ {\bf{9550635}}$  & $ {\bf{9550635}}$   & $ {\bf{119339}}$ & $ {\bf{35627}}$ & $ {\bf{35627}}$ & $ {\bf{-781}}$ & $ {\bf{3003}}$ & $ {\bf{3003}}$ & $ {\bf{6}}$ & $ {\bf{6}}$ & $ {\bf{2827}}$ & $ {\bf{1163}}$ & $ {\bf{43}}$ & $ {\bf{27}}$ & $ {\bf{210}}$ & $ {\bf{210}}$\\ \hline
\rule{0pt}{3ex} \rule[-1ex]{0pt}{0pt} {$7$} & $ {\bf{63532485}}$ & $ {\bf{63532485}}$   & $ {\bf{-468027}}$ & $ {\bf{134597}}$ & $ {\bf{134597}}$ & $ {\bf{-4811}}$ & $ {\bf{8073}}$ & $ {\bf{8073}}$ & $ {\bf{135}}$ & $ {\bf{135}}$ & $ {\bf{-7547}}$ & $ {\bf{2437}}$ & $ {\bf{197}}$ & $ {\bf{1}}$ & $ {\bf{385}}$ & $ {\bf{385}}$\\ \hline
\rule{0pt}{3ex} \rule[-1ex]{0pt}{0pt} {$8$} & $ {\bf{347643114}}$ & $ {\bf{347643114}}$   & $ {\bf{1511146}}$ & $ {\bf{392426}}$ & $ {\bf{392426}}$ & $ {\bf{1274}}$ & $ {\bf{16224}}$ & $ {\bf{16224}}$ & $ {\bf{-219}}$ & $ {\bf{-219}}$ & $ {\bf{14378}}$ & $ {\bf{4394}}$ & $ {\bf{-5341}}$ & $ {\bf{78}}$ & $ {\bf{539}}$ & $ {\bf{539}}$\\ \hline
\rule{0pt}{3ex} \rule[-1ex]{0pt}{0pt} {$9$} & $ {\bf{356054375}}$  & $ {\bf{356054375}}$  & $ {\bf{-1901977}}$ & $ {\bf{419175}}$ & $ {\bf{419175}}$ & $ {\bf{-1377}}$ & $ {\bf{18227}}$ & $ {\bf{18227}}$ & $ {\bf{-160}}$ & $ {\bf{-160}}$ & $ {\bf{-18425}}$ & $ {\bf{8327}}$ & $ {\bf{-153}}$ & $ {\bf{27}}$ & $ {\bf{650}}$ & $ {\bf{650}}$\\ \hline
\rule{0pt}{3ex} \rule[-1ex]{0pt}{0pt} {$10$} & $ {\bf{1407126890}}$  & $ {\bf{1407126890}}$  & $ {\bf{-3199638}}$ & $ {\bf{789866}}$ & $ {\bf{789866}}$ & $ {\bf{25194}}$ & $ {\bf{25103}}$ & $ {\bf{25103}}$ & $ {\bf{155}}$ & $ {\bf{155}}$ & $ {\bf{-18326}}$ & $ {\bf{3178}}$ & $ {\bf{-406}}$ & $ {\bf{26}}$ & $ {\bf{615}}$ & $ {\bf{615}}$\\ \hline
\rule{0pt}{3ex} \rule[-1ex]{0pt}{0pt} {$11$} & $ {\bf{3214743741}}$  & $ {\bf{3214743741}}$  & $ {\bf{3059133}}$ & $ {\bf{1214653}}$ & $ {\bf{1214653}}$ & $ {\bf{22933}}$ & $ {\bf{32670}}$ & $ {\bf{32670}}$ & $ {\bf{270}}$ & $ {\bf{270}}$ & $ {\bf{17501}}$ & $ {\bf{-3875}}$ & $ {\bf{1469}}$ & $ {\bf{-27}}$ & $ {\bf{616}}$ & $ {\bf{616}}$\\ \hline
\rule{0pt}{3ex} \rule[-1ex]{0pt}{0pt} {$12$} & $ {\bf{4221380670}}$  & $ {\bf{4221380670}}$   & $ {\bf{-999362}}$ & $ {\bf{1615934}}$ & $ {\bf{1615934}}$ & $ {\bf{-44642}}$ & $ {\bf{37401}}$ & $ {\bf{37401}}$ & $ {\bf{465}}$ & $ {\bf{465}}$ & $ {\bf{-8130}}$ & $ {\bf{-8130}}$ & $ {\bf{-706}}$ & $ {\bf{-78}}$ & $ {\bf{595}}$ & $ {\bf{595}}$\\ \hline
\rule{0pt}{3ex} \rule[-1ex]{0pt}{0pt} {$13$} & $ {\bf{4275362520}}$  & $ {\bf{4275362520}}$   & $ {\bf{10237656}}$ & $ {\bf{1710808}}$ & $ {\bf{1710808}}$ & $ {\bf{48568}}$ & $ {\bf{42471}}$ & $ {\bf{42471}}$ & $ {\bf{594}}$ & $ {\bf{594}}$ & $ {\bf{45144}}$ & $ {\bf{20056}}$ & $ {\bf{1240}}$ & $ {\bf{0}}$ & $ {\bf{770}}$ & $ {\bf{770}}$\\ \hline
\rule{0pt}{3ex} \rule[-1ex]{0pt}{0pt} {$14$} & $ {\bf{4622913750}}$  & $ {\bf{4622913750}}$  & $ {\bf{11656918}}$ & $ {\bf{2011350}}$ & $ {\bf{2011350}}$ & $ {\bf{-42042}}$ & $ {\bf{58422}}$ & $ {\bf{58422}}$ & $ {\bf{345}}$ & $ {\bf{345}}$ & $ {\bf{57750}}$ & $ {\bf{22678}}$ & $ {\bf{-1066}}$ & $ {\bf{-78}}$ & $ {\bf{1275}}$ & $ {\bf{1275}}$\\ \hline
\rule{0pt}{3ex} \rule[-1ex]{0pt}{0pt} {$15$} & $ {\bf{9287037474}}$  & $ {\bf{9287037474}}$   & $ {\bf{16720418}}$ & $ {\bf{1896994}}$ & $ {\bf{1896994}}$ & $ {\bf{87074}}$ & $ {\bf{19734}}$ & $ {\bf{19734}}$ & $ {\bf{1023}}$ & $ {\bf{1023}}$ & $ {\bf{24354}}$ & $ {\bf{31522}}$ & $ {\bf{2850}}$ & $ {\bf{26}}$ & $ {\bf{99}}$ & $ {\bf{99}}$\\ \hline
\rule{0pt}{3ex} \rule[-1ex]{0pt}{0pt} {$16$} & $ {\bf{12501781215}}$  & $ {\bf{12501781215}}$  & $ {\bf{-15693601}}$ & $ {\bf{2075359}}$ & $ {\bf{2075359}}$ & $ {\bf{65807}}$ & $ {\bf{52404}}$ & $ {\bf{52404}}$ & $ {\bf{1293}}$ & $ {\bf{1293}}$ & $ {\bf{-47201}}$ & $ {\bf{11935}}$ & $ {\bf{-1057}}$ & $ {\bf{-53}}$ & $ {\bf{715}}$ & $ {\bf{715}}$\\ \hline
\rule{0pt}{3ex} \rule[-1ex]{0pt}{0pt} {$17$} & $ {\bf{13508418144}}$  & $ {\bf{13508418144}}$    & $ {\bf{-18677152}}$ & $ {\bf{3512928}}$ & $ {\bf{3512928}}$ & $ {\bf{-134368}}$ & $ {\bf{57135}}$ & $ {\bf{57135}}$ & $ {\bf{1488}}$ & $ {\bf{1488}}$ & $ {\bf{-60576}}$ & $ {\bf{18272}}$ & $ {\bf{3680}}$ & $ {\bf{0}}$ & $ {\bf{694}}$ & $ {\bf{694}}$\\ \hline
\rule{0pt}{3ex} \rule[-1ex]{0pt}{0pt} {$18$} & $ {\bf{27416186875}}$  & $ {\bf{27416186875}}$   & $ {\bf{-37543429}}$ & $ {\bf{6369275}}$ & $ {\bf{6369275}}$ & $ {\bf{-44149}}$ & $ {\bf{129349}}$ & $ {\bf{129349}}$ & $ {\bf{-1385}}$ & $ {\bf{-1385}}$ & $ {\bf{-125125}}$ & $ {\bf{32827}}$ & $ {\bf{2299}}$ & $ {\bf{-1}}$ & $ {\bf{1925}}$ & $ {\bf{1925}}$\\ \hline
\rule{0pt}{3ex} \rule[-1ex]{0pt}{0pt} {$19$} & $ {\bf{75844139371}}$  & $ {\bf{75844139371}}$  & $ {\bf{-56581525}}$ & $ {\bf{3402091}}$ & $ {\bf{3402091}}$ & $ {\bf{162435}}$ & $ {\bf{-1001}}$ & $ {\bf{-1001}}$ & $ {\bf{2887}}$ & $ {\bf{2887}}$ & $ {\bf{4235}}$ & $ {\bf{49931}}$ & $ {\bf{875}}$ & $ {\bf{-273}}$ & $ {\bf{-154}}$ & $ {\bf{-154}}$\\ \hline
\rule{0pt}{3ex} \rule[-1ex]{0pt}{0pt} {$20$} & $ {\bf{80426400000}}$  & $ {\bf{80426400000}}$   & $ {\bf{-87975680}}$ & $ {\bf{10451200}}$ & $ {\bf{10451200}}$ & $ {\bf{104960}}$ & $ {\bf{169170}}$ & $ {\bf{169170}}$ & $ {\bf{-120}}$ & $ {\bf{-120}}$ & $ {\bf{-172800}}$ & $ {\bf{77056}}$ & $ {\bf{-1792}}$ & $ {\bf{0}}$ & $ {\bf{1750}}$ & $ {\bf{1750}}$\\ \hline
\rule{0pt}{3ex} \rule[-1ex]{0pt}{0pt} {$21$} & $ {\bf{90807234375}}$  & $ {\bf{90807234375}}$   & $ {\bf{55087175}}$ & $ {\bf{2498375}}$ & $ {\bf{2498375}}$ & $ {\bf{-118625}}$ & $ {\bf{-10725}}$ & $ {\bf{-10725}}$ & $ {\bf{-1815}}$ & $ {\bf{-1815}}$ & $ {\bf{-9625}}$ & $ {\bf{42215}}$ & $ {\bf{-3001}}$ & $ {\bf{-325}}$ & $ {\bf{0}}$ & $ {\bf{0}}$\\ \hline
\rule{0pt}{3ex} \rule[-1ex]{0pt}{0pt} {$22$} & $ {\bf{90807234375}}$  & $ {\bf{90807234375}}$    & $ {\bf{55087175}}$ & $ {\bf{2498375}}$ & $ {\bf{2498375}}$ & $ {\bf{-118625}}$ & $ {\bf{-10725}}$ & $ {\bf{-10725}}$ & $ {\bf{-1815}}$ & $ {\bf{-1815}}$ & $ {\bf{-9625}}$ & $ {\bf{42215}}$ & $ {\bf{-3001}}$ & $ {\bf{-325}}$ & $ {\bf{0}}$ & $ {\bf{0}}$\\ \hline
\rule{0pt}{3ex} \rule[-1ex]{0pt}{0pt} {$\cdots $} & $ {\cdots}$ & $ {\cdots}$   & $ {\cdots}$ & $ {\cdots}$ & $ {\cdots}$ & $ {\cdots}$ & $ {\cdots}$ & $ {\cdots}$ & $ {\cdots}$ & $ {\cdots}$ & $ {\cdots}$ & $ {\cdots}$ & $ {\cdots}$ & $ {\cdots}$ & $ {\cdots}$ & $ {\cdots}$\\ \hline

\rule{0pt}{3ex} \rule[-1ex]{0pt}{0pt} {$185$} & $ {\bf{96256}}$  & $ {\bf{-96256}}$  & $ {\bf{0}}$ & $ {\bf{2048}}$ & $ {\bf{-2048}}$ & $ {\bf{0}}$ & $ {\bf{352}}$ & $ {\bf{-352}}$ & $ {\bf{28}}$ & $ {\bf{-28}}$ & $ {\bf{0}}$ & $ {\bf{0}}$ & $ {\bf{0}}$ & $ {\bf{0}}$ & $ {\bf{56}}$ & $ {\bf{-56}}$\\ \hline
\rule{0pt}{3ex} \rule[-1ex]{0pt}{0pt} {$186$} & $ {\bf{10506240}}$  & $ {\bf{-10506240}}$   & $ {\bf{0}}$ & $ {\bf{45056}}$ & $ {\bf{-45056}}$ & $ {\bf{0}}$ & $ {\bf{3456}}$ & $ {\bf{-3456}}$ & $ {\bf{-108}}$ & $ {\bf{108}}$ & $ {\bf{0}}$ & $ {\bf{0}}$ & $ {\bf{0}}$ & $ {\bf{0}}$ & $ {\bf{240}}$ & $ {\bf{-240}}$\\ \hline
\rule{0pt}{3ex} \rule[-1ex]{0pt}{0pt} {$187$} & $ {\bf{410132480}}$  & $ {\bf{-410132480}}$   & $ {\bf{0}}$ & $ {\bf{516096}}$ & $ {\bf{-516096}}$ & $ {\bf{0}}$ & $ {\bf{23648}}$ & $ {\bf{-23648}}$ & $ {\bf{-4}}$ & $ {\bf{4}}$ & $ {\bf{0}}$ & $ {\bf{0}}$ & $ {\bf{0}}$ & $ {\bf{0}}$ & $ {\bf{880}}$ & $ {\bf{-880}}$\\ \hline
\rule{0pt}{3ex} \rule[-1ex]{0pt}{0pt} {$188$} & $ {\bf{8844386304}}$  & $ {\bf{-8844386304}}$    & $ {\bf{0}}$ & $ {\bf{3629056}}$ & $ {\bf{-3629056}}$ & $ {\bf{0}}$ & $ {\bf{96096}}$ & $ {\bf{-96096}}$ & $ {\bf{840}}$ & $ {\bf{-840}}$ & $ {\bf{0}}$ & $ {\bf{0}}$ & $ {\bf{0}}$ & $ {\bf{0}}$ & $ {\bf{1904}}$ & $ {\bf{-1904}}$\\ \hline
\rule{0pt}{3ex} \rule[-1ex]{0pt}{0pt} {$189$} & $ {\bf{36657653760}}$ & $ {\bf{-36657653760}}$   & $ {\bf{0}}$ & $ {\bf{8198144}}$ & $ {\bf{-8198144}}$ & $ {\bf{0}}$ & $ {\bf{146016}}$ & $ {\bf{-146016}}$ & $ {\bf{-432}}$ & $ {\bf{432}}$ & $ {\bf{0}}$ & $ {\bf{0}}$ & $ {\bf{0}}$ & $ {\bf{0}}$ & $ {\bf{1960}}$ & $ {\bf{-1960}}$\\ \hline
\rule{0pt}{3ex} \rule[-1ex]{0pt}{0pt} {$190$} & $ {\bf{53936390144}}$ & $ {\bf{-53936390144}}$  & $ {\bf{0}}$ & $ {\bf{-1835008}}$ & $ {\bf{1835008}}$ & $ {\bf{0}}$ & $ {\bf{11648}}$ & $ {\bf{-11648}}$ & $ {\bf{1280}}$ & $ {\bf{-1280}}$ & $ {\bf{0}}$ & $ {\bf{0}}$ & $ {\bf{0}}$ & $ {\bf{0}}$ & $ {\bf{-56}}$ & $ {\bf{56}}$\\ \hline
\rule{0pt}{3ex} \rule[-1ex]{0pt}{0pt} {$191$} & $ {\bf{53936390144}}$ & $ {\bf{-53936390144}}$   & $ {\bf{0}}$ & $ {\bf{-1835008}}$ & $ {\bf{1835008}}$ & $ {\bf{0}}$ & $ {\bf{11648}}$ & $ {\bf{-11648}}$ & $ {\bf{1280}}$ & $ {\bf{-1280}}$ & $ {\bf{0}}$ & $ {\bf{0}}$ & $ {\bf{0}}$ & $ {\bf{0}}$ & $ {\bf{-56}}$ & $ {\bf{56}}$\\ \hline
\rule{0pt}{3ex} \rule[-1ex]{0pt}{0pt} {$\cdots $} & $ {\cdots}$ & $ {\cdots}$   & $ {\cdots}$ & $ {\cdots}$ & $ {\cdots}$ & $ {\cdots}$ & $ {\cdots}$ & $ {\cdots}$ & $ {\cdots}$ & $ {\cdots}$ & $ {\cdots}$ & $ {\cdots}$ & $ {\cdots}$ & $ {\cdots}$ & $ {\cdots}$ & $ {\cdots}$\\ \hline
\end{tabular}
}
\caption{Partial character table of the baby Monster group $2 \cdot \mathbb{B}$.}
\end{table}
\end{landscape}

\begin{landscape}
\begin{table}[h!]
\resizebox{23.0cm}{!}{
\centering
\renewcommand{\arraystretch}{1.1}
\begin{tabular}{|c||c|c|c|c|c|c|c|c|c|c|c|c|c|c|c|c|c|c|c|}
\hline
 {$n$}  & 1A & 3A & 3B & 2A & 6A & 6B &2B & 6C & 6D & 3C & 3D & 3E & 3F & 3G & 3H & 3I & 4A & 12A & 12B \\ \hline \hline
\rule{0pt}{3ex} \rule[-1ex]{0pt}{0pt}  {$1$} & $ {\bf{1}}$       & $ {\bf{1}}$        & $ {\bf{1}}$       & $ {\bf{1}}$     & $ {\bf{1}}$     & $ {\bf{1}}$   & $ {\bf{1}}$     & $ {\bf{1}}$      & $ {\bf{1}}$       & $ {\bf{1}}$       & $ {\bf{1}}$      & $ {\bf{1}}$     & $ {\bf{1}}$    & $ {\bf{1}}$   & $ {\bf{1}}$    & $ {\bf{1}}$     & $ {\bf{1}}$   & $ {\bf{1}}$   & $ {\bf{1}}$ \\ \hline 
\rule{0pt}{3ex} \rule[-1ex]{0pt}{0pt} {$2$} & $ {\bf{8671}}$    & $ {\bf{8671}}$     & $ {\bf{8671}}$    & $ {\bf{351}}$   & $ {\bf{351}}$   & $ {\bf{351}}$  & $ {\bf{-33}}$   & $ {\bf{-33}}$    & $ {\bf{-33}}$     & $ {\bf{247}}$     & $ {\bf{-77}}$    & $ {\bf{-77}}$   & $ {\bf{-77}}$ & $ {\bf{85}}$  & $ {\bf{85}}$   & $ {\bf{85}}$    & $ {\bf{31}}$  & $ {\bf{31}}$  & $ {\bf{31}}$\\ \hline
\rule{0pt}{3ex} \rule[-1ex]{0pt}{0pt} {$3$} & $ {\bf{57477}}$   & $ {\bf{57477}}$    & $ {\bf{57477}}$   & $ {\bf{1157}}$  & $ {\bf{1157}}$  & $ {\bf{1157}}$ & $ {\bf{133}}$   & $ {\bf{133}}$    & $ {\bf{133}}$     & $ {\bf{534}}$     & $ {\bf{615}}$    & $ {\bf{615}}$   & $ {\bf{615}}$ & $ {\bf{210}}$ & $ {\bf{210}}$  & $ {\bf{210}}$   & $ {\bf{69}}$  & $ {\bf{69}}$  & $ {\bf{69}}$ \\ \hline
\rule{0pt}{3ex} \rule[-1ex]{0pt}{0pt} {$4$} & $ {\bf{249458}}$  & $ {\bf{249458}}$   & $ {\bf{249458}}$  & $ {\bf{2354}}$  & $ {\bf{2354}}$  & $ {\bf{2354}}$ & $ {\bf{370}}$   & $ {\bf{370}}$    & $ {\bf{370}}$     & $ {\bf{2705}}$    & $ {\bf{869}}$    & $ {\bf{869}}$   & $ {\bf{869}}$ & $ {\bf{167}}$ & $ {\bf{167}}$  & $ {\bf{167}}$   & $ {\bf{50}}$  & $ {\bf{50}}$  & $ {\bf{50}}$ \\ \hline
\rule{0pt}{3ex} \rule[-1ex]{0pt}{0pt} {$5$} & $ {\bf{555611}}$  & $ {\bf{555611}}$   & $ {\bf{555611}}$  & $ {\bf{5083}}$  & $ {\bf{5083}}$  & $ {\bf{5083}}$ & $ {\bf{91}}$    & $ {\bf{91}}$     & $ {\bf{91}}$      & $ {\bf{-535}}$    & $ {\bf{2300}}$   & $ {\bf{2300}}$  & $ {\bf{2300}}$ & $ {\bf{518}}$ & $ {\bf{518}}$  & $ {\bf{518}}$   & $ {\bf{155}}$  & $ {\bf{155}}$  & $ {\bf{155}}$  \\ \hline
\rule{0pt}{3ex} \rule[-1ex]{0pt}{0pt} {$6$} & $ {\bf{1603525}}$ & $ {\bf{1603525}}$  & $ {\bf{1603525}}$ & $ {\bf{1925}}$  & $ {\bf{1925}}$  & $ {\bf{1925}}$ & $ {\bf{-315}}$  & $ {\bf{-315}}$   & $ {\bf{-315}}$    & $ {\bf{6475}}$    & $ {\bf{-275}}$   & $ {\bf{-275}}$  & $ {\bf{-275}}$ & $ {\bf{-140}}$ & $ {\bf{-140}}$ & $ {\bf{-140}}$ & $ {\bf{5}}$  & $ {\bf{5}}$  & $ {\bf{5}}$ \\ \hline
\rule{0pt}{3ex} \rule[-1ex]{0pt}{0pt} {$7$} & $ {\bf{1603525}}$ & $ {\bf{1603525}}$  & $ {\bf{1603525}}$ & $ {\bf{1925}}$  & $ {\bf{1925}}$  & $ {\bf{1925}}$ & $ {\bf{-315}}$  & $ {\bf{-315}}$   & $ {\bf{-315}}$    & $ {\bf{6475}}$    & $ {\bf{-275}}$   & $ {\bf{-275}}$  & $ {\bf{-275}}$ & $ {\bf{-140}}$ & $ {\bf{-140}}$ & $ {\bf{-140}}$ & $ {\bf{5}}$ & $ {\bf{5}}$  & $ {\bf{5}}$\\ \hline
\rule{0pt}{3ex} \rule[-1ex]{0pt}{0pt} {$8$} & $ {\bf{1666833}}$ & $ {\bf{1666833}}$  & $ {\bf{1666833}}$ & $ {\bf{11153}}$ & $ {\bf{11153}}$ & $ {\bf{11153}}$ & $ {\bf{273}}$   & $ {\bf{273}}$    & $ {\bf{273}}$     & $ {\bf{3093}}$    & $ {\bf{-1848}}$  & $ {\bf{-1848}}$ & $ {\bf{-1848}}$ & $ {\bf{987}}$  & $ {\bf{987}}$ & $ {\bf{987}}$ & $ {\bf{209}}$  & $ {\bf{209}}$  & $ {\bf{209}}$\\ \hline
\rule{0pt}{3ex} \rule[-1ex]{0pt}{0pt} {$9$} & $ {\bf{4864431}}$ & $ {\bf{4864431}}$  & $ {\bf{4864431}}$ & $ {\bf{13871}}$ & $ {\bf{13871}}$ & $ {\bf{13871}}$ & $ {\bf{-1105}}$ & $ {\bf{-1105}}$  & $ {\bf{-1105}}$   & $ {\bf{4836}}$    & $ {\bf{4917}}$   & $ {\bf{4917}}$   & $ {\bf{4917}}$ & $ {\bf{867}}$  & $ {\bf{867}}$ & $ {\bf{867}}$ & $ {\bf{175}}$  & $ {\bf{175}}$  & $ {\bf{175}}$\\ \hline
\rule{0pt}{3ex} \rule[-1ex]{0pt}{0pt} {$\cdots$} & $ \cdots$ & $ \cdots$ & $ \cdots$ & $ \cdots$ & $ \cdots$ & $ \cdots$ & $ \cdots$ & $ \cdots$ & $ \cdots$ & $ \cdots$ & $ \cdots$ & $ \cdots$ & $ \cdots$ & $ \cdots$ & $ \cdots$ & $ \cdots$ & $ \cdots$ & $ \cdots$ & $ \cdots$ \\ \hline
\rule{0pt}{3ex} \rule[-1ex]{0pt}{0pt} {$109$} & $ {\bf{783}}$      & $ {\bf{783 \alpha}}$             & $ {\bf{783 \overline{\alpha}}}$  & $ {\bf{79}}$    & $ {\bf{79 \alpha}}$             & $ {79 \bf{\overline{\alpha}}}$ & $ {\bf{15}}$  & $ {\bf{15 \alpha}}$             & $ {\bf{15 \overline{\alpha}}}$   & $ {\bf{0}}$  & $ {\bf{54}}$  & $ {\bf{54 \alpha}}$             & $ {\bf{54 \overline{\alpha}}}$   & $ {\bf{27}}$  & $ {\bf{27 \alpha}}$             & $ {\bf{27\overline{\alpha}}}$  & $ {\bf{15}}$  & $ {\bf{15 \alpha}}$             & $ {\bf{15\overline{\alpha}}}$\\ \hline
\rule{0pt}{3ex} \rule[-1ex]{0pt}{0pt} {$110$} & $ {\bf{783}}$      & $ {\bf{783\overline{\alpha}}}$  & $ {783\bf{\alpha}}$             & $ {\bf{79}}$    & $ {\bf{79\overline{\alpha}}}$  & $ {79\bf{\alpha}}$            & $ {\bf{15}}$  & $ {\bf{15\overline{\alpha}}}$             & $ {\bf{15\alpha}}$              & $ {\bf{0}}$    & $ {\bf{54}}$  & $ {\bf{54\overline{\alpha}}}$             & $ {\bf{54\alpha}}$ & $ {\bf{27}}$  & $ {\bf{27\overline{\alpha}}}$             & $ {\bf{27\alpha}}$ & $ {\bf{15}}$  & $ {\bf{15\overline{\alpha}}}$             & $ {\bf{15\alpha}}$\\ \hline
\rule{0pt}{3ex} \rule[-1ex]{0pt}{0pt} {$111$} & $ {\bf{64584}}$    & $ {\bf{64584\alpha}}$             & $ {\bf{64584\overline{\alpha}}}$  & $ {\bf{1352}}$  & $ {\bf{1352\alpha}}$             & $ {\bf{1352\overline{\alpha}}}$ & $ {\bf{72}}$  & $ {\bf{72 \alpha}}$             & $ {\bf{72 \overline{\alpha}}}$   & $ {\bf{0}}$    & $ {\bf{-297}}$  & $ {\bf{-297\alpha}}$             & $ {\bf{-297\overline{\alpha}}}$ & $ {\bf{216}}$  & $ {\bf{216 \alpha}}$             & $ {\bf{216 \overline{\alpha}}}$ & $ {\bf{72}}$  & $ {\bf{72 \alpha}}$             & $ {\bf{72 \overline{\alpha}}}$\\ \hline  
\rule{0pt}{3ex} \rule[-1ex]{0pt}{0pt} {$112$} & $ {\bf{64584}}$    & $ {\bf{64584\overline{\alpha}}}$  & $ {\bf{64584 \alpha}}$             & $ {\bf{1352}}$  & $ {\bf{1352\overline{\alpha}}}$  & $ {\bf{1352\alpha}}$            & $ {\bf{72}}$  & $ {\bf{72\overline{\alpha}}}$             & $ {72\bf{\alpha}}$              & $ {\bf{0}}$    & $ {\bf{-297}}$  & $ {\bf{-297\overline{\alpha}}}$             & $ {\bf{-297\alpha}}$  & $ {\bf{216}}$  & $ {\bf{216\overline{\alpha}}}$             & $ {\bf{216 \alpha}}$ & $ {\bf{72}}$  & $ {\bf{72\overline{\alpha}}}$             & $ {72\bf{\alpha}}$\\ \hline
\rule{0pt}{3ex} \rule[-1ex]{0pt}{0pt} {$113$} & $ {\bf{306513}}$   & $ {\bf{306513 \alpha}}$             & $ {\bf{306513 \overline{\alpha}}}$  & $ {\bf{3433}}$  & $ {\bf{3433 \alpha}}$             & $ {\bf{3433\overline{\alpha}}}$ & $ {\bf{489}}$  & $ {\bf{489 \alpha}}$             & $ {\bf{489 \overline{\alpha}}}$   & $ {\bf{0}}$    & $ {\bf{1431}}$  & $ {\bf{1431 \alpha}}$             & $ {\bf{1431\overline{\alpha}}}$  & $ {\bf{351}}$  & $ {\bf{351 \alpha}}$             & $ {\bf{351\overline{\alpha}}}$ & $ {\bf{105}}$  & $ {\bf{105 \alpha}}$             & $ {\bf{105\overline{\alpha}}}$\\ \hline
\rule{0pt}{3ex} \rule[-1ex]{0pt}{0pt} {$114$} & $ {\bf{306513}}$   & $ {\bf{306513\overline{\alpha}}}$  & $ {\bf{306513 \alpha}}$             & $ {\bf{3433}}$  & $ {\bf{3433\overline{\alpha}}}$  & $ {\bf{3433 \alpha}}$            & $ {\bf{489}}$  & $ {\bf{489\overline{\alpha}}}$             & $ {\bf{489 \alpha}}$              & $ {\bf{0}}$    & $ {\bf{1431}}$  & $ {\bf{1431\overline{\alpha}}}$             & $ {\bf{1431 \alpha}}$  & $ {\bf{351}}$  & $ {\bf{351\overline{\alpha}}}$             & $ {\bf{351\alpha}}$ & $ {\bf{105}}$  & $ {\bf{105\overline{\alpha}}}$             & $ {\bf{105 \alpha}}$\\ \hline
\rule{0pt}{3ex} \rule[-1ex]{0pt}{0pt} {$115$} & $ {\bf{306513}}$   & $ {\bf{306513 \alpha}}$             & $ {\bf{306513\overline{\alpha}}}$  & $ {\bf{2729}}$  & $ {\bf{2729 \alpha}}$             & $ {\bf{2729\overline{\alpha}}}$ & $ {\bf{-279}}$  & $ {\bf{-279\alpha}}$             & $ {\bf{-279\overline{\alpha}}}$   & $ {\bf{0}}$    & $ {\bf{1431}}$  & $ {\bf{1431 \alpha}}$             & $ {\bf{1431\overline{\alpha}}}$  & $ {\bf{351}}$  & $ {\bf{351 \alpha}}$             & $ {\bf{351\overline{\alpha}}}$ & $ {\bf{105}}$  & $ {\bf{105\alpha}}$             & $ {\bf{105\overline{\alpha}}}$\\ \hline 
\rule{0pt}{3ex} \rule[-1ex]{0pt}{0pt} {$116$} & $ {\bf{306513}}$   & $ {\bf{306513\overline{\alpha}}}$  & $ {\bf{306513 \alpha}}$             & $ {\bf{2729}}$  & $ {\bf{2729\overline{\alpha}}}$  & $ {\bf{2729 \alpha}}$            & $ {\bf{-279}}$  & $ {\bf{-279\overline{\alpha}}}$             & $ {\bf{-279 \alpha}}$              & $ {\bf{0}}$    & $ {\bf{1431}}$  & $ {\bf{1431\overline{\alpha}}}$             & $ {\bf{1431 \alpha}}$  & $ {\bf{351}}$  & $ {\bf{351\overline{\alpha}}}$             & $ {\bf{351 \alpha}}$ & $ {\bf{105}}$  & $ {\bf{105\overline{\alpha}}}$             & $ {\bf{105 \alpha}}$\\ \hline
\rule{0pt}{3ex} \rule[-1ex]{0pt}{0pt} {$117$} & $ {\bf{6724809}}$  & $ {\bf{6724809 \alpha}}$             & $ {\bf{6724809\overline{\alpha}}}$  & $ {\bf{26377}}$ & $ {\bf{26377 \alpha}}$             & $ {\bf{26377\overline{\alpha}}}$ & $ {\bf{-567}}$  & $ {\bf{-567 \alpha}}$             & $ {\bf{-567\overline{\alpha}}}$   & $ {\bf{0}}$    & $ {\bf{-3861}}$  & $ {\bf{-3861 \alpha}}$             & $ {\bf{-3861\overline{\alpha}}}$  & $ {\bf{2079}}$  & $ {\bf{2079 \alpha}}$             & $ {\bf{2079\overline{\alpha}}}$ & $ {\bf{393}}$  & $ {\bf{393 \alpha}}$             & $ {\bf{393\overline{\alpha}}}$ \\ \hline
\rule{0pt}{3ex} \rule[-1ex]{0pt}{0pt} {$118$} & $ {\bf{6724809}}$  & $ {\bf{6724809\overline{\alpha}}}$  & $ {\bf{6724809 \alpha}}$             & $ {\bf{26377}}$ & $ {\bf{26377\overline{\alpha}}}$  & $ {\bf{26377 \alpha}}$            & $ {\bf{-567}}$  & $ {\bf{-567\overline{\alpha}}}$             & $ {\bf{-567 \alpha}}$              & $ {\bf{0}}$    & $ {\bf{-3861}}$  & $ {\bf{-3861\overline{\alpha}}}$             & $ {\bf{-3861\alpha}}$  & $ {\bf{2079}}$  & $ {\bf{2079\overline{\alpha}}}$             & $ {\bf{2079 \alpha}}$ & $ {\bf{393}}$  & $ {\bf{393\overline{\alpha}}}$             & $ {\bf{393 \alpha}}$\\ \hline
\rule{0pt}{3ex} \rule[-1ex]{0pt}{0pt} {$\cdots$} & $ \cdots$ & $ \cdots$ & $ \cdots$ & $ \cdots$ & $ \cdots$ & $ \cdots$ & $ \cdots$ & $ \cdots$ & $ \cdots$ & $ \cdots$ & $ \cdots$ & $ \cdots$ & $ \cdots$ & $ \cdots$ & $ \cdots$ & $ \cdots$ & $ \cdots$ & $ \cdots$ & $ \cdots$ \\ \hline
\end{tabular}
}
\caption{Partial character table of the Fischer group $3 \cdot \text{Fi}_{24}'$. Here $\alpha = \frac{-1+i\sqrt{3}}{2}$ and $\overline{\alpha} = \frac{-1-i\sqrt{3}}{2}$.}
\end{table}
\end{landscape}

\section{Generalized Bilinear Relations } \label{App:C}

\subsection{$2 \cdot \mathbb{B}$}
We find the twined characters of baby Monster CFT for various $g \in 2 \cdot \mathbb{B}$ which are combined with the characters of
Ising models and form the Mckay-Thompson series for various $g_{\mathbb{M}} \in \mathbb{M}$. More preciesly, general expression
 of generalized bilinear relation have a form of
\begin{align}
\begin{split}
  j^{g_{\mathbb{M}}}(\tau) &= f_0(\tau) \cdot \tilde f^{g}_0(\t) + f_\e(\t) \cdot \tilde f^{g}_\e(\tau) + f_\s(\t) \cdot \tilde f^{g}_\s(\tau),
\end{split}
\end{align}
where $\tilde f^{g}(\tau)$ is twined character for $g \in 2 \cdot \mathbb{B}$ and $j^{g_{\mathbb{M}}}(\tau)$ denotes Mckay-Thompson 
series for class $g_{\mathbb{M}} \in \mathbb{M}$. Below table present which twined character forming  Mckay-Thompson series of type 
$g_{\mathbb{M}}$. 
For instance, once we have twined characters of baby Monster CFT for $g=2C$,
they merge with the characters of Ising model to produce Mckay-Thompson series of class $2A$.
\begin{table}[h]
\centering
\renewcommand{\arraystretch}{1.1}
\begin{tabular}{|c|c||c|c|| c|c|}
\hline
 $g$ & $g_{\mathbb{M}}$ &  $g$ & $g_{\mathbb{M}}$  &  $g$ & $g_{\mathbb{M}}$ \\ \hline \hline
{2A} & \rule{0pt}{2.8ex}$ { 2A }$  & 3B & 3B & 6A & 6A \\ \hline
{2B} & \rule{0pt}{2.8ex}$ { 2A }$  & 4A &  4B & 6B & 6D \\ \hline
{2C} & \rule{0pt}{2.8ex}$ { 2A }$ & 4B & 4A & 8A & 8C \\ \hline
{2D} & \rule{0pt}{2.8ex}${ 2B }$ & 4D & 4C & 5A & 5A \\ \hline
{3A} & \rule{0pt}{2.8ex}$ { 3A }$ & 4C & 4A & 10A & 10A \\ \hline
\end{tabular}
\caption{Generalized bilinear relation for $2 \cdot \mathbb{B}$}
\label{External Bilinear for 2B}
\end{table}

\subsection{$3 \cdot \text{Fi}_{24}'$}
It turns out that the twined characters of $c=\frac{116}{5}$ putative CFT for various $g \in 3 \cdot \text{Fi}_{24}'$ 
also constitute the Mckay-Thompson series of certain class $g_{\mathbb{M}} \in \mathbb{M}$ with the characters of three-states Potts model. We find that 
the explicit form of the generalized bilinear relation is given by
\begin{align}
\begin{split}
  j^{g_{\mathbb{M}}}(\tau) &= f_0(\tau) \cdot \tilde f^{g}_0(\t) + f_1(\t) \cdot \tilde f^{g}_1(\tau) +  f_2(\t) \cdot \tilde f^{g}_2(\tau)  + f'_2(\t) \cdot \tilde f{'}^{g}_2(\tau) \\
                                           &\quad +  f_3(\t) \cdot \tilde f^{g}_3(\tau) +  f'_3(\t) \cdot \tilde f{'}^{g}_3(\tau),
\end{split}
\end{align}
where $\tilde f^{g}(\tau)$ is twined character for $g \in 3 \cdot \text{Fi}_{24}'$ and $j^{g_{\mathbb{M}}}(\tau)$ denotes Mckay-Thompson 
series for class $g_{\mathbb{M}} \in \mathbb{M}$ as before. Below table exhibit which twined character yields Mckay-Thompson 
series of certain class, like table \ref{External Bilinear for 2B} describes.
\begin{table}[h]
\centering
\renewcommand{\arraystretch}{1.1}
\begin{tabular}{|c|c||c|c|| c|c|}
\hline
 $g$ & $g_{\mathbb{M}}$ &  $g$ & $g_{\mathbb{M}}$  &  $g$ & $g_{\mathbb{M}}$ \\ \hline \hline
{3A,3B} & \rule{0pt}{2.8ex}$ { 3A }$  & 6C,6D &  6C & 3G  & 3A  \\ \hline
{2A} & \rule{0pt}{2.8ex}$ { 2A }$  & 3C &  3A  & 3H,3I  &  3B  \\ \hline
{6A,6B} & \rule{0pt}{2.8ex}$ { 6A }$ & 3D & 3B  & 4A &  4A  \\ \hline
{2B} & \rule{0pt}{2.8ex}${ 2B }$ & 3E,3F & 3A  & 12A,12B &  12A  \\ \hline
\end{tabular}
\caption{Generalized bilinear relation for $3 \cdot \text{Fi}_{24}'$}
\end{table}



\bibliographystyle{jhep}
\bibliography{refs_MA}

\end{document}